\documentclass{aa}

\usepackage{natbib,epsfig,amsmath,amssymb}

\def\ga{\,\hbox{\hbox{$ > $}\kern -0.8em \lower 1.0ex\hbox{$\sim$}}\,}
\def\la{\,\hbox{\hbox{$ < $}\kern -0.8em \lower 1.0ex\hbox{$\sim$}}\,}
\def\beq{\begin{equation}}
\def\eeq{\end{equation}}

\newcommand\bb[1] {   \mbox{\boldmath{$#1$}}  }

\newcommand\bcdot{\bb{\cdot}}
\newcommand\btimes{\bb{\times}}

\titlerunning{Simulations of galactic disk}
\authorrunning{Hennebelle \& Iffrig}

\begin{document}

\title{Simulations of magnetized multiphase galactic disk regulated by supernovae explosions}

\author{Patrick Hennebelle\inst{1,2}, Olivier Iffrig\inst{1}}

\institute{
Laboratoire AIM, 
Paris-Saclay, CEA/IRFU/SAp - CNRS - Universit\'e Paris Diderot, 91191, 
Gif-sur-Yvette Cedex, France \\
\and
LERMA (UMR CNRS 8112), Ecole Normale Sup\'erieure, 75231 Paris Cedex, France}

\abstract{What exactly controls star formation in the Galaxy remains controversial. In particular, the role of 
feedback and magnetic field are still partially understood.} {We investigate the role played by supernovae feedback and
magnetic field onto the star formation and the structure of the Galactic disk. } {We perform
numerical simulations of the turbulent, magnetized, self-gravitating, multi-phase, supernovae regulated ISM within
a 1 kpc stratified box. 
We implemented various schemes for the supernovae. This goes from a random distribution at a fixed rate to 
 distributions for which the supernovae are spatially and temporally correlated to the formation of stars. 
To study the influence of  magnetic field on  star formation, we perform both  hydrodynamical and magneto-hydrodynamical simulations. 
  } {We find that supernovae feedback  has a drastic influence on the galactic evolution. The supernovae 
distribution is playing a very significant role. When the supernovae are not correlated with  star formation 
events, they do not modify significantly the very high  star formation rate obtained without feedback. When the supernovae follow
the accretion, the star formation rate can be reduced by a factor up to 30. Magnetic field is also playing a significant role. It reduces 
the star formation rate by a factor up to 2-3 and reduces the number of collapse sites by a factor of about 2.  
} {The exact correlation between the  supernovae and the dense gas appears to have significant 
consequences on the galactic disk evolution and the star formation. This implies that small scale studies 
are necessary to understand and quantify the feedback efficiency. Magnetic field does influence the 
star formation at galactic scales by reducing the star formation rate and the number of star formation sites.} 

\keywords{magnetoydrodynamics (MHD) --   Instabilities  --  Interstellar  medium:
kinematics and dynamics -- structure -- clouds -- Star: formation}

\maketitle

\section{Introduction}

Star formation is a multi-scale and multi-physics problem, which is only partially 
understood. In particular what physical process is responsible for the 
relatively low star formation of the Milky-way \citep[e.g.][]{zuckerman+1974, dobbs+2013}
remains a subject of controversy. 

Historically three main physical processes have been emphasized, namely
magnetic field \citep[e.g.][]{shu+1987}, turbulence \citep[e.g.][]{maclow+2004},
and stellar feedback \citep[e.g.][]{maclow2013,agertz+2013}  which
includes supernovae explosions, ionising radiation, heating by stellar radiation, stellar outflows and stellar winds.
 While magnetic field 
has been measured to have substantial intensities \citep{crutcher2012}, it may 
be nevertheless too weak to reduce the star formation rate (SFR) by orders of magnitude. 
The effects of turbulence and feedback are not straighforward to disantangle.
In particular, turbulence decays in a few crossing times \citep[e.g.][]{maclow+2004} 
and must be fed at large scales  either through galactic 
large scale spiral waves or  by the various sources 
of stellar feedback. The feedback may therefore play a dual role in limiting the amount
of mass that is eventually accreted by stars while in the same time triggering 
large scale turbulence. In any case, previous studies, which have been simulating a whole galaxy found a clear 
impact of the feedback onto the SFR
\citep[e.g.][]{tasker-bryan2006,dubois+2008,bournaud+2010,kim+2011,dobbs+2011,tasker2011,hopkins+2011,renaud+2013}, 
which is able to reduce  star formation substantially, may be up to observed values. 
Although there is a general agreement that the simulations without feedback present too high an SFR, 
the amount by which it is reduced when feedback is introduced depends on which feedback is introduced and how.
For example \citet{tasker-bryan2006} who perfomed simulations with supernovae feedback found that the 
SFR is reduced by a factor of about 2, \citet{tasker2011} included the UV radiation feedback found that 
the SFR is typically reduced by a factor of 1.5-2. Finally, \citet{hopkins+2011} have introduced the radiative feedback 
assuming that the radiation of stars can efficienly communicate its momentum to the gas. They found that 
the SFR can be reduced by a factor on the order of  10 to 30.

In spite of these studies, the exact roles played by feedback, both for triggering the turbulence and for 
limiting the star formation, as well as by magnetic field are only partially understood. 
First of all, given the limited resolution of large scale studies (typically a few pc) the exact way 
feedback is applied remains partially arbitrary. In particular, 
the first pioneering  studies which have been first focussing on kpc scales
regulated by supernovae explosions \citep{deavillez+2005,joung+2006}, did not 
include self-gravity and therefore could not associate supernovae with star formation events accurately
 though they unanbigously show the relevance of these studies. 
Even when self-gravity is included, the exact influence of the choices made to determine their locations 
has not been explored clearly. Second of all, the influence that the magnetic field has on the star formation 
rate at the kpc scale is  less explored. The only simulations that include both self-gravity and magnetic 
field which have been performed to date are presented in  \citet{Wang+2009,rudiger+2013}.  These authors have been concluding that 
magnetic field reduces the star formation rate by a factor of a few. Moreover these 2 studies modeled a whole galaxy 
implying that the spatial resolution is necessary limited to describe the ISM structure.

In parallel to the numerical studies \citep{slyz+2005,deavillez+2005,joung+2006,hill+2012, kim+2011, kim+2013, gent+2013},
a few analytical models have been developed and compared with observations and simulations 
\citep{ostriker+2010,kim+2011,faucher+2013}. For example, in their model, \citet{ostriker+2010}
 consider vertical mechanical equilibrium between gravity and pressure 
(mainly thermal and kinetic) as well as thermal equilibrium in the ISM (i.e. equilibrium between heating from stars and cooling).
It is then further assumed that the thermal and turbulent supports are proportional. This leads them to predict the SFR as 
a function of the column density through the galactic planes and the model
compares well with a sample of observations \citep{leroy+2008} 
and simulations \citep{kim+2011}. In their models, \citet{faucher+2013} consider a galactic disk, which 
is also in equilibrium along the vertical direction but assume that the galactic disk has a Toomre parameter 
\citep{toomre1964} of about $Q \simeq 1$,
 that is to say is in marginal equilibrium and self-regulates. They then perform an energy budget between the 
energy dissipated by units of time through turbulent cascade and the energy injected through stellar feedback. 
In both cases important assumptions are made regarding  how energy and momentum are injected within the dense gas. 
While it is clearly unavoidable  to make such assumptions, given the difficulty of the problem, it 
is nevertheless important to understand how  exactly are momentum and energy injected, more precisely how 
they  distribute between diffuse and dense gas. More generally,   what are the uncertainties induced by our 
incomplete understanding of the correlation between massive stars, at the origin of most of the feedback, and the 
surrounding dense material ? It is the purpose of the present paper to address these issues.

In this paper  we adopt a similar  setup  to the one adopted by 
 \citet{slyz+2005,deavillez+2005,joung+2006,hill+2012, kim+2011, kim+2013, gent+2013}, 
that is to say a  kpc  simulation of a galactic disk in which turbulence is driven 
by supernovae remnants.
Previous studies have been finding that they can reproduce many of the 
interstellar medium feature such as multi-phase ISM, approximate energy equipartition between thermal, magnetic and kinetic 
energies, galactic outflows and formation of molecular clouds therefore demonstating the interest of performing
this type of simulations. Indeed, this range of scales is a good compromise between the need for enough 
numerical resolution to describe sufficiently well the ISM physics and the amount of computational power
 available on present computers.  
While the results obtained previously are encouraging, only few of these works have been treating self-gravity and none 
of these works have been considering magnetic field and self-gravity, which is mandory for a proper physical description.

 We include both self-gravity and magnetic field and  we explore various schemes for the supernovae feedback 
going from a random distribution to a distribution in which supernovae are correlated both spatially and temporally with star formation.
This spatial and temporal correlation turns out to be drastically important when self-gravity is self-consistently treated. 
The primary reason is that when a dense region undergoes gravitationnal collapse, it becomes largely decoupled 
from the surrounding medium and therefore little influenced by the supernovae which may be exploding nearby. 
It is only if a supernova explodes 
within the collapsing region and while collapsing is occuring that feedback has a significant impact and can 
reduce the mass that is eventually accreted.

The second section of the paper presents the numerical setup and in particular discusses
the various schemes we have been developing to implement supernovae feedback. 
In the third section we describe the disk structure in the various models while 
in fourth section we investigate the properties of the multi-phase ISM in two of 
the simulations. In section 5 we discuss the star formation rate and the mass distribution 
of the sink particles formed in the different simulations. Section six concludes the paper.

\section{Numerical setup}

\setlength{\unitlength}{1cm}
\begin{figure*} 
\begin{picture} (0,18)
\put(0,8){\includegraphics[width=9cm]{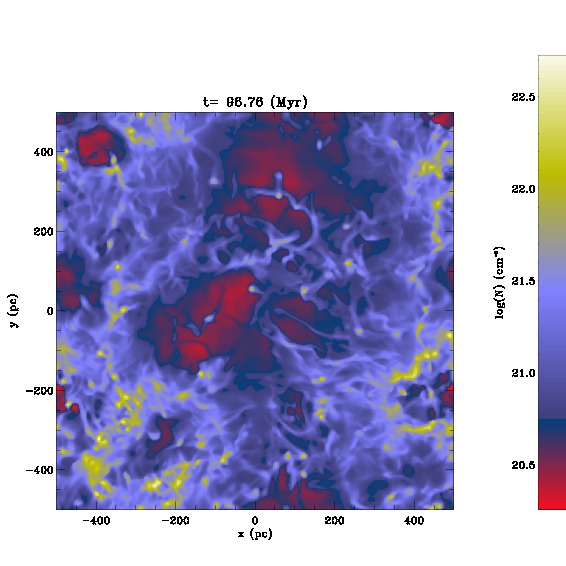}} 
\put(9,8){\includegraphics[width=9cm]{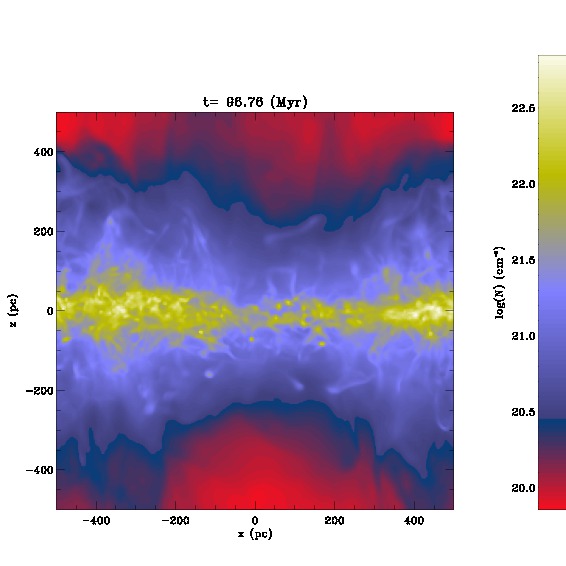}}
\put(0,0){\includegraphics[width=9cm]{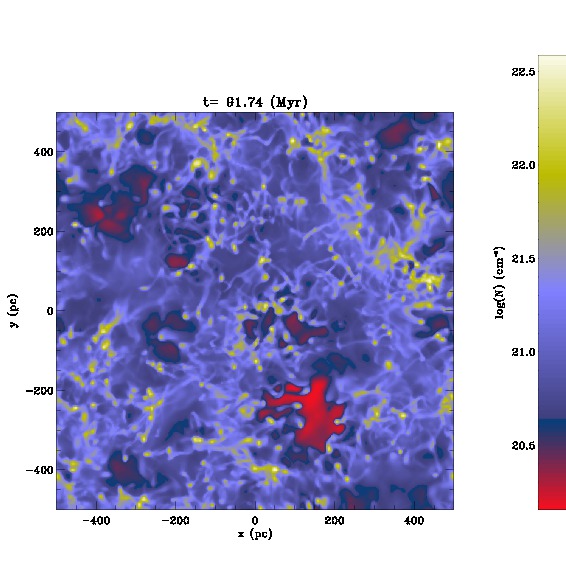}}  
\put(9,0){\includegraphics[width=9cm]{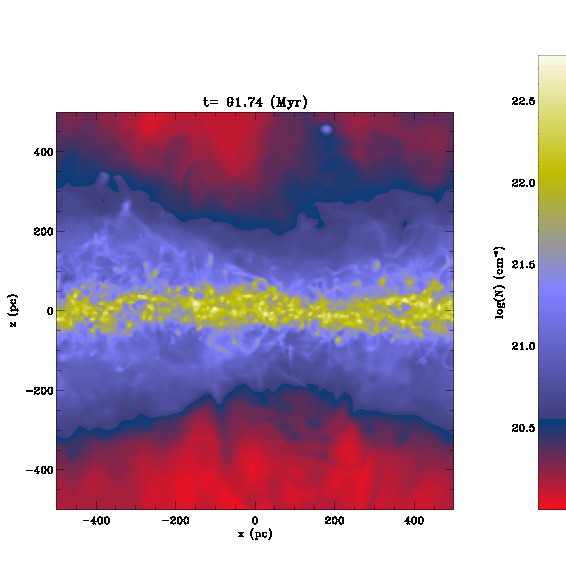}}  
\end{picture}
\caption{Column density along the $z$-axis (left panels) and along the $y$-axis (right panels)
for  MHD run C1 (upper panels) and hyrodynamical run C2 (lower panels). 
 The left panels illustrate the complex multi-phase structure of the galactic plane
while the right panels show the stratification induced by the gravitational field of the galaxy. 
 In the hydrodynamical run, the interstellar medium is more fragmented than in the MHD one.}
\label{image}
\end{figure*}

\setlength{\unitlength}{1cm}
\begin{figure*} 
\begin{picture} (0,18)
\put(0,0){\includegraphics[width=9cm]{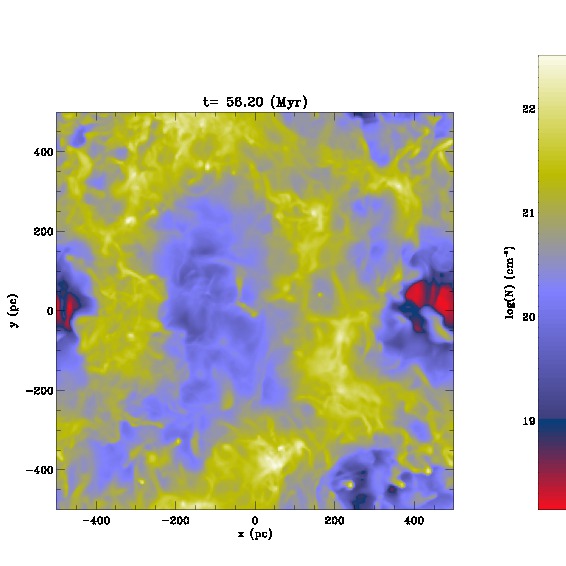}} 
\put(9,0){\includegraphics[width=9cm]{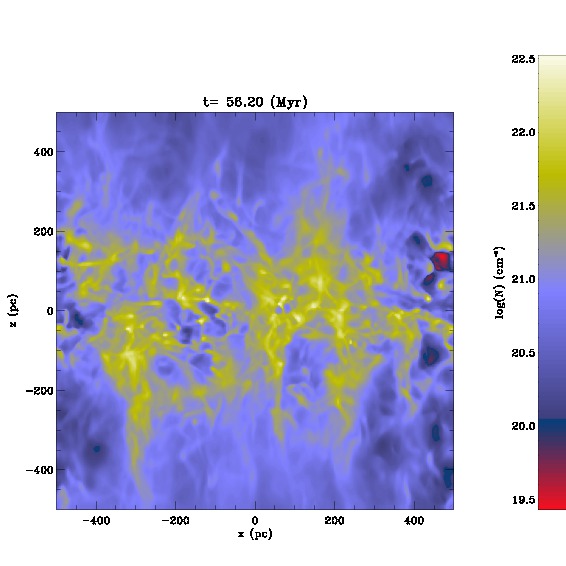}}
\put(0,8){\includegraphics[width=9cm]{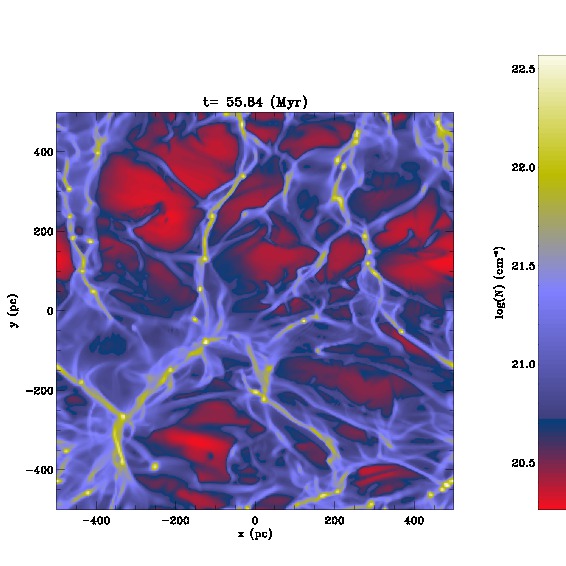}}  
\put(9,8){\includegraphics[width=9cm]{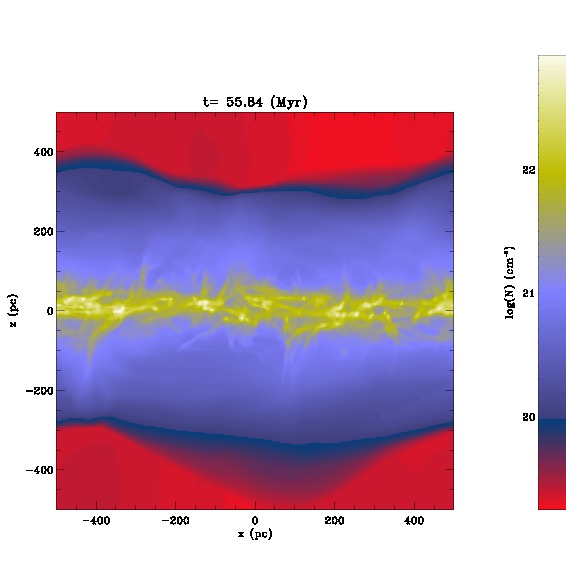}}  
\end{picture}
\caption{Column density along the $z$-axis (left panels) and along the $y$-axis (right panels)
for the run with no feedback, NF1, (upper panels) and the run with scheme D (lower panels).
Comparison with Fig.~\ref{image} clearly shows the drastic impact of the feedback.  In particular, 
the run NF1 shows thin and long filaments in which self-gravitating fragments develop, indicating that 
the gas is primarily organised by self-gravity and undergoes run away collapse. The galactic disk is considerably thinner
than when supernovae are included. On the other hand, run D shows a broader 
galactic disk than in runs C1 and C2 illustrating the importance of the spatial correlation between 
supernovae explosions and dense gas.} 
\label{image2}
\end{figure*}

\subsection{Physical processes and initial conditions}
The physical processes and initial conditions are similar to what has been 
described by previous authors \citep[e.g.][]{deavillez+2005,joung+2006, hill+2012, kim+2013}.
We consider a 1 kpc computational box in which a gravitational field is applied along the 
$z$-axis. Its value is identical to the choice made by \citet{joung+2006} 
taken from \citet{Kuijken+1989}
and is given by 
\begin{eqnarray}
g(z) = - { a_1 z \over \sqrt{z^2+z_0^2} } -a_2 z,
\label{grav_pot}
\end{eqnarray}
where $a_1=1.42 \times 10^{-3}$ kpc Myr$^{-2}$, $a_2= 5.49 \times 10^{-4}$ Myr$^{-2}$
and $z_0=0.18$ kpc. This gravitational field represents the contribution 
of the stars and dark matter in our Galaxy. \\

We solve the ideal magneto-hydrodynamical (MHD) equations in the presence of self and external gravity
and include cooling and heating processes relevant for the ISM. The equations 
are given by 
\begin{eqnarray}
\frac{\partial \rho}{\partial t} + \bb{\nabla} \bcdot (\rho \bb{v})
&=& 0, \\
\rho \left[ \frac{\partial \bb{v}}{\partial t} + (\bb{v} \bcdot
  \bb{\nabla}) \bb{v} \right] &=& -\bb{\nabla} P  +  
\frac{(\bb{\nabla} \btimes \bb{B}) \btimes \bb{B}}{4\pi} \nonumber \\
&& - \rho {\bf g} - \rho \bb{\nabla} \phi, \label{impulsion_eq}\\ 
\rho \left[ \frac{\partial \bb{e}}{\partial t} + (\bb{v} \bcdot
  \bb{\nabla}) \bb{e} \right] &=& - P (\bb{\nabla} \bcdot
\bb{v}) - \rho {\cal L}, \label{internal_nrj_eq} \\
\frac{\partial \bb{B}}{\partial t} &=& \bb{\nabla} \btimes (\bb{v}
\btimes \bb{B}), \label{induction_eq} \\
\Delta \phi &=& 4 \pi G \rho, \label{poisson}
\end{eqnarray}
where all notation have their usual meaning. 
The heating and cooling terms, which appear in the loss function ${\cal L}$,
 are identical to the one considered in 
\citet{ah05}, i.e. include UV heating  due to photoelectric effect on grains, 
Lyman-$\alpha$, oxygen and ionised carbon cooling as well as cooling due to the 
recombination onto grains \citep[see e.g.][]{wolfire+2003}. 
Note that at this stage we use a constant UV field which is not correlated 
to the star formation rate \citep[e.g.][]{tasker2011,kim+2013}.
At temperature larger than 10$^4$ K, we use the
fit provided in \citet{joung+2006}  for the cooling function inferred by
\citet{sutherland+1993}.  
Coriolis and centrifugal forces are not included  at this stage.

At the beginning of the simulation, the density distribution is 
given by 
\begin{eqnarray}
n(z) = n_0 \exp \left(- \left( { z \over z_0} \right)^2 \right), 
\label{dens0}
\end{eqnarray}
with $n_0=1.5$ cm$^{-3}$ and $z_0=150$ pc. 
 The temperature is initially equal to about 8000 K which corresponds 
to the temperature of the warm neutral gas (WNM). 
A {\it turbulent} velocity field 
is generated using random phase and a Kolmogorov powerspectrum. Its rms 
amplitude is equal to 5 km s$^{-1}$. Finally, the magnetic field is initially 
aligned along the $x$-axis and is proportional to the density field. Its value
in the equatorial plane is about 2.5$\mu$G for our fiducial run (later named run C1), the 
value of 0.5$\mu$G is also explored.

\subsection{Supernovae prescriptions}
To take into account the supernovae feedback we first select a position as described below.
Then we increase the thermal energy of all the cells located at a distance smaller than 
three grid cells from the supernova center in such a way that the thermal energy is uniform 
in this sphere and equal to $10^{51}$ erg. In most, but not all, of our runs we have also 
introduced a kinetic feedback. This is achieved by adding to the corresponding cells a
radial  homologous velocity field (proportional to the distance from the supernova center).
The total kinetic energy is equal to 5\% of the thermal energy which corresponds to 
the typical  momentum that is injected at the end of the Sedov Phase,  that is to say during the phase for which the supernova
expansion remains nearly adiabatic. Indeed when the shell surrounding the supernova bubble becomes radiative, most of the energy 
is radiated away and the expansion proceeds at constant momentum \cite[e.g.][]{chevalier77}. Fortunately 
enough this momentum is largely independent of the density field \citep{blondin+1998} even when
it is highly irregular (Iffrig \& Hennebelle 2014, in prep).
Note that in principle, the momentum should be self-consistently generated during the 
Sedov expansion. However, the lack of resolution does not guarantee that this phase is well treated 
when the supernovae explodes in a dense region. In any case, as described below we have 
also performed the case without kinetic feedback for comparison. These two runs probably 
constitute  upper and lower limits. 

The spatial location of the supernovae is another important aspect. Previous 
authors \citep[e.g.][]{deavillez+2005,joung+2006} have been 
distributing them randomly or in correlation with density. These authors also tested the case where 
the supernovae are clustered \citep[see e.g.][ for a description]{joung+2006}. 
In this work,  four spatial and temporal supernovae distributions  have been tested, hereafter 
scheme A, B, C and D. 
We recall that an important difference with these classical studies is that self-gravity 
is treated. 

Scheme A is very similar to the scheme described for example in \citet{joung+2006}. 
The supernovae are distributed randomly in the 
$x$ and $y$-directions. To mimic the observed supernovae distribution in the 
Milky Way, their z-coordinate follows a Gaussian distribution of thickness
equal to 150 pc. Their rate is equal to the observed galactic rate and is equal to 
1/50 per year.  One difference is that we use a fixed radius for the supernovae
remnant rather than using a radius which enclosed a fixed mass. This implies 
that the timesteps in the simulation can be quite low since the temperature 
is higher when a supernova explodes in a diffuse medium. 
Another difference 
is that we do not redistribute the mass within the supernova radius as 
\citet{joung+2006} who impose a uniform gas density inside the supernova bubbles. 
In principle, it could be worth testing all these choices but here we focus on a 
different issue.

Scheme B consists looking at the density maximum  in the simulations and  
choosing its  position for the supernova center. The rate is also equal to 
the galactic one.  A supernovae is introduced only if the density peak is larger 
than $10$ cm$^{-3}$ in the simulation. However,  denser gas develops rapidly in the simulation
and except for the very first, the supernovae are generally associated to gas 
of densities $10^2-10^3$ cm$^{-3}$ which is present at all time. This scheme has the advantage to have 
a good spatial correlation with star formation events. However it does not have any temporal correlation
since the supernovae rate is fixed and equal to the galactic one.

Scheme C and D are different and take advantage of the sink particles used in our simulations. 
Each time a sink particle has accreted 120 solar masses, we place a supernova in its neighborhood. 
This prescription is motivated by the typical abundance of stars more massive than 
the 8 solar masses needed to give raise to a  supernova explosion.  However, we do not place the 
supernova center directly at the sink position for various reasons. First of all, as described 
later the sink particles have a radius of  4 computational cells, corresponding 
to  16 pc with our current resolution. This number of cells for the sink radius is typical to what is
usually assumed \citep[e.g.][]{Krumholz+04}. By definition it sets the limit of the resolution in the 
simulation and inside the sink, the gas distribution is not well described. Second of all, 
it takes at least 4 Myr for the most massive stars to explode. During this time both the star and 
the cloud have evolved. For example in 10 Myr, a star which moves at a velocity of 1 km s$^{-1}$, 
will have cover a distance of about 10 pc. Finally, because of the other sources of feedback, 
massive stars may have pushed the dense gas away before supernovae explode.
To test the importance of the spatial correlation between supernovae and sink particles, 
we have implemented two prescriptions. First (scheme C), we place randomly the supernovae 
within the sink particles radius. Second (Scheme D), we place them within a shell 
of inner radius equal to the sink radius and outer radius equal to two times this value. 
As  will be seen later, these two schemes lead to similar but not identical results.

Another important issue is the time delay that should be also taken into account since 
supernovae typically explode between about 4 and 40 Myr after the formation of the massive star.
Although we note that some delay is introduced with our scheme since, at least 120 solar masses
of gas have to be accreted before supernovae take place, it is in most of the time shorter (10$^5-10^6$ yr)
than a few Myr. 
However as emphasized in other studies \citep[e.g.][]{matzner2002, dale+2012, dale+2013, agertz+2013, kim+2013},
other sources of feedback, namely ionising radiation, winds and jets, start influencing the surrounding clouds much earlier. 
These sources should be taken into account as well. Since we feel it is important to go step by step, we postpone 
studying these effects and concentrate for now on supernovae only.

\subsection{Numerical code and resolution}
\label{code}
To carry out our numerical simulations, we employed RAMSES \citep{teyssier2002, fromang+2006}, 
an adaptive mesh refinement (AMR) code that uses Godunov schemes to solve the MHD equations
and the constrained transport method to ensure that ${\rm div}B$ is maintained 
at zero within machine accuracy. For most runs, we do not use the AMR capacity  and 
keep the resolution fix using 256 grid points in each direction. 
However, we also perform a run with one more AMR level introduced when the density 
reaches a value of 10 cm$^{-3}$.
 The computational 
box size is equal to 1 kpc and the spatial resolution is 4 pc. This choice is dictated 
by the very large number of time steps ($\simeq$ 50,000-100,000), which are required for these calculations.
This is because  supernovae feedback introduced very high velocities of the order of 
few 100 km s$^{-1}$ as well as temperature as high as $10^8$ K in few cells. Moreover, we integrate far enough in order to make 
sure that some quasi-stationary regime has been reached. This will be assessed by verifying that 
the mean profile of various  quantities such as densities does not vary significantly with time. It should be stressed however that 
 since self-gravity is treated and accretion is occuring onto the sink particles, no strict stationarity can be reached. 
 
We use periodic boundary conditions in $x$ and $y$ directions and outflow
condition at the $z$ boundaries. In particular, this implies that the gas ejected from the galactic 
disk can escape the computational box.

In this work sink particles (implemented in the public version of RAMSES) are being 
used to follow the dense regions which have collapsed under the influence of self-gravity. 
They closely follow the implementation of \citet{Krumholz+04}.
The sinks are introduced when a density threshold of 10$^3$ cm$^{-3}$ is reached. Their radius is equal to
4 grid cells. A new sink can be created only if it is not located closer than 10 grid cells from 
another sink. When  the sinks get too close, i.e. closer than one grid cell, 
they get  merge using a friend of friend procedure. Finally, the sinks accrete gas from 
surrounding cells if they are located at less than a sink radius and if the density is larger than 
10$^3$ cm$^{-3}$. The sink particles interact with the surrounding medium through the gravitational field.
The contribution of the sink to the gravitational potential 
 is included using a particle-mesh approach, that is to say the mass within the sink is projected onto
the grid and added to  the gas density when the Poisson equation is solved.

\subsection{Comparison between various setups}
It is worth emphazising the differences between the various setups which have been used so far, keeping in mind 
that  given the complexity of the problem under investigation, $i)$ it is hard to include every relevant process, 
$ii)$ it is important to perform studies which make different choices and approximations to disantangle 
the effects of the different physical processes.

In the setup used by  \citet{deavillez+2005, joung+2006} and 
\citet{hill+2012}, hydro or MHD equations are solved using a 500 pc or a 1 kpc size box
in the equatorial plane but a much larger scale height (typically 5 to 10 times these values).
This insures a good description of the galactic wind and the disk-halo connection though 
large scale modes may be filtered out by the elongated box. 
\citet{gent+2013} proceed somewhat similarly but also include the galactic shear in their study.
In these studies the supernovae explode randomly or are correlated with density peaks. 
None of these studies include self-gravity.

\citet{kim+2011, kim+2013} do include self-gravity and like us, consider a cubic computational box.
These choices are clearly putting more emphasis on the disk itself and therefore on the 
star formation, which is taking place,  than on the galactic outflows and the halo. There is an important 
difference  regarding the forcing. They estimate the number of massive stars that should form 
in a given region rather than following the accretion onto sink particles and they consider 
only the mechanical feedback from supernovae. 

As will be seen below, the different prescriptions leads to quite different results. 
In particular, the correlation between star forming gas and the feedback is a necessary condition
to prevent  very efficient star formation. This is however true mostly if self-gravity is 
included. 

\subsection{Runs performed}
In the present paper, we perform various runs to test the influence that magnetic 
field has onto the galactic disk evolution and to study the influence of 
the various prescriptions for the supernovae feedback. 
In our fiducial run (later referred as  C1), the magnetic intensity has an intensity in the equatorial 
plane of about 2.5 $\mu$G and scheme C is used for the supernovae with both 
thermal and kinetic feedback. 

To study the influence of the magnetic field, we perform an hydrodynamical run 
(later referred as C2) and a run with a lower magnetisation
initially equal to about 0.5 $\mu$G in the midplane (run C3). Apart from the strength of the field, 
these runs are identical to run C1. 

To study the influence of the feedback scheme, we perform a series of runs identical 
to run C1 apart for the feedback scheme. First, we run 
two cases without any feedback, one purely hydrodynamical (NF2) and one with 2.5 $\mu$G initially (NF1). Second, we consider two cases 
with scheme A and B respectively (simply labelled runs A and B) and both thermal and kinetic feedback. Third 
we perform a simulation with supernovae scheme C but with thermal feedback only (run C4). Fourth we 
carry out a calculation with scheme D (run D). 

Finally, to investigate the important issue of numerical resolution, we also present 
a run identical to run C1 but using another level of refinement leading to an effective 
resolution of about 2 pc. The refinement is performed when the cell density 
reaches a threshold of 10 cm$^{-3}$ leading to a total number cells at this level comparable to the 
number of cells in the non-refined runs. The results are presented in the Appendix.

Table~\ref{run} summarizes the runs performed in the paper
and provide consistent label.

\begin{table}
\begin{tabular} {| {l} | {l} | {l} | {l} | {l} |  }
\hline
 label   & physics & scheme & feedback & resolution \\
\hline
 NF1   & MHD (2.5 $\mu$G) & - & - & 256$^3$ \\
\hline
 NF2   & hydro & - & - & 256$^3$ \\
\hline
 A   & MHD (2.5 $\mu$G) & A & thermal+kinetic & 256$^3$  \\
\hline
 B   & MHD (2.5 $\mu$G) & B & thermal+kinetic & 256$^3$ \\
\hline
 C1   & MHD (2.5 $\mu$G) & C & thermal+kinetic & 256$^3$  \\
\hline
 C1b   & MHD (2.5 $\mu$G) & C & thermal+kinetic & 512$^3$  \\
\hline
 C2   & hydro & C & thermal+kinetic & 256$^3$ \\
\hline
 C3   & MHD (0.5 $\mu$G) & C & thermal+kinetic & 256$^3$ \\
\hline
 C4   & MHD (2.5 $\mu$G) & C & thermal & 256$^3$ \\
\hline
 D   & MHD (2.5 $\mu$G) & D & thermal+kinetic  & 256$^3$ \\
\hline
\end{tabular}
\caption{Summary of the different runs performed in the paper.
The scheme refers to the way supernovae are being introduced 
in the simulations. Schemes A and B assume a constant supernova rates and are not correlated 
with sink particles. With schemes C and D the supernovae are correlated spatially and temporally
with sink particles (see text).}
\label{run}
\end{table}

\setlength{\unitlength}{1cm}
\begin{figure} 
\begin{picture} (0,23.)
\put(0,18){\includegraphics[width=8cm]{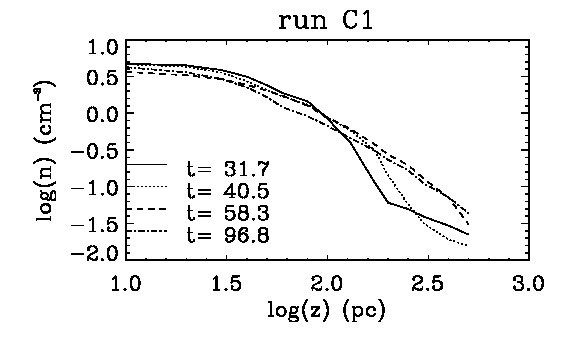}} 
\put(0,0){\includegraphics[width=8cm]{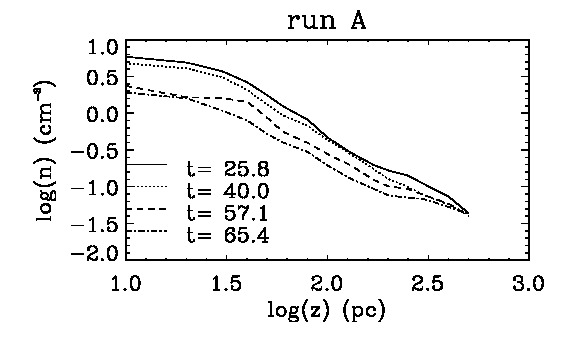}} 
\put(0,4.5){\includegraphics[width=8cm]{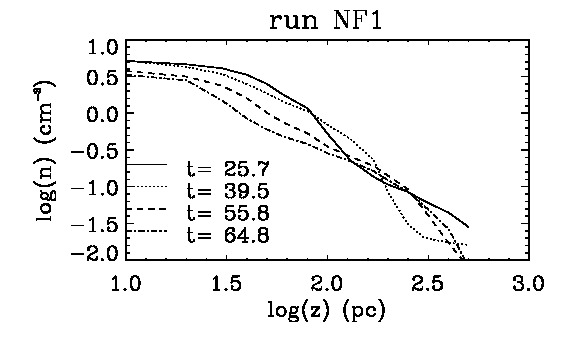}} 
\put(0,9){\includegraphics[width=8cm]{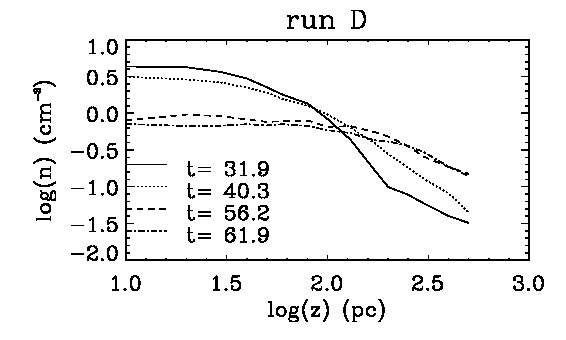}} 
\put(0,13.5){\includegraphics[width=8cm]{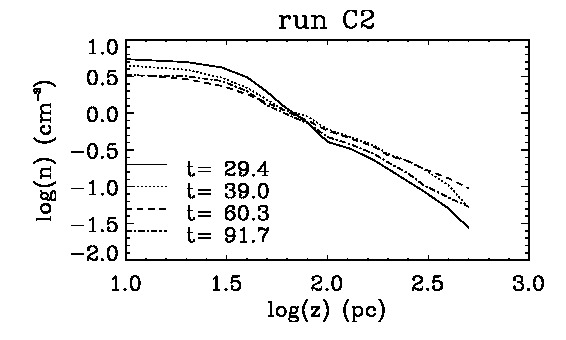}} 
\end{picture}
\caption{Mean density profile along the $z$-axis for five different models
(see label) at four different timesteps. The disk profile is 
much thinner for runs NF1 and A than for runs C1, C2 and D.}
\label{rho_z}
\end{figure}

\setlength{\unitlength}{1cm}
\begin{figure} 
\begin{picture} (0,23.)
\put(0,18){\includegraphics[width=8cm]{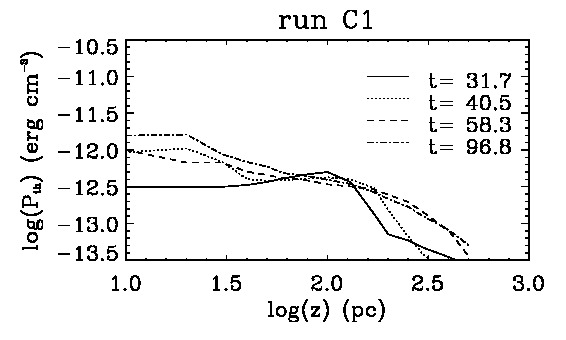}} 
\put(0,0){\includegraphics[width=8cm]{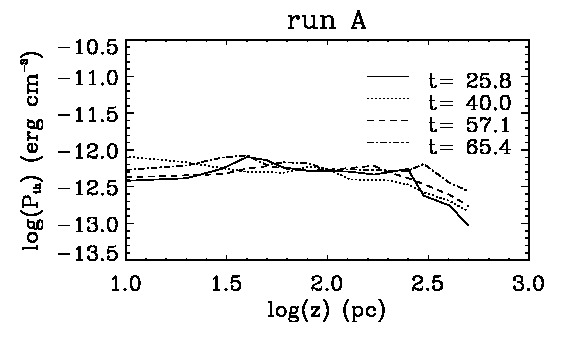}} 
\put(0,4.5){\includegraphics[width=8cm]{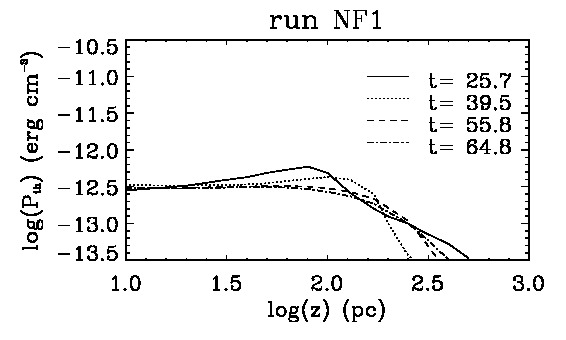}} 
\put(0,9){\includegraphics[width=8cm]{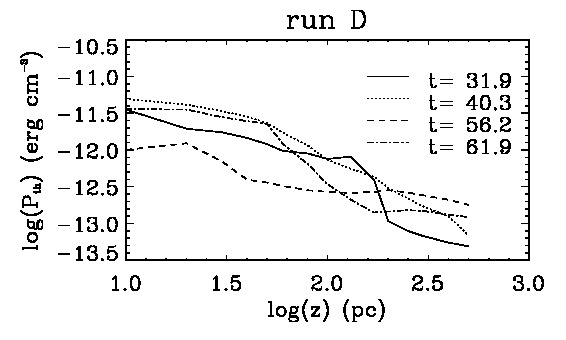}} 
\put(0,13.5){\includegraphics[width=8cm]{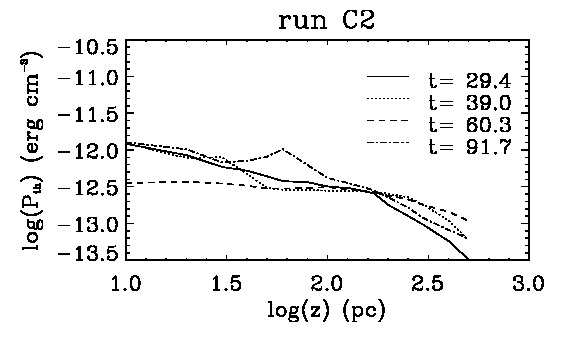}} 
\end{picture}
\caption{Thermal pressure profile along the $z$-axis for five different models
(see label) at four different timesteps. The largest values are obtained 
for run D and the smallest for run NF1 which has no feedback.} 
\label{Pth_z}
\end{figure}

\setlength{\unitlength}{1cm}
\begin{figure} 
\begin{picture} (0,23.)
\put(0,18){\includegraphics[width=8cm]{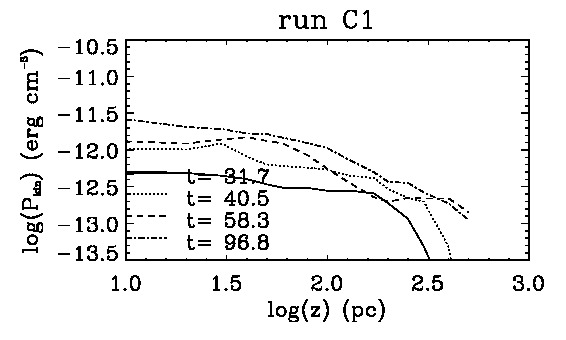}} 
\put(0,0){\includegraphics[width=8cm]{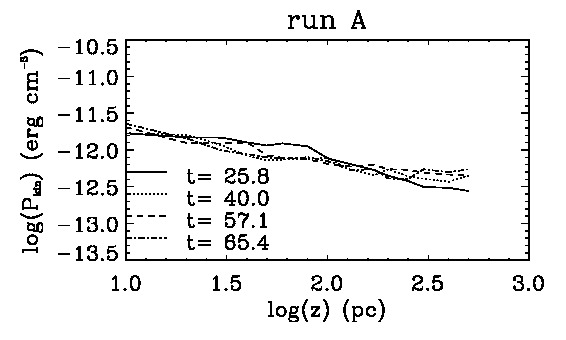}} 
\put(0,4.5){\includegraphics[width=8cm]{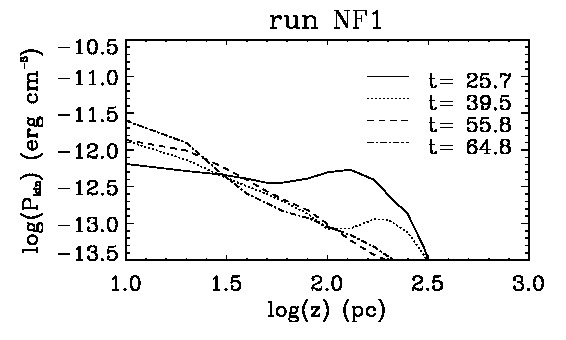}} 
\put(0,9){\includegraphics[width=8cm]{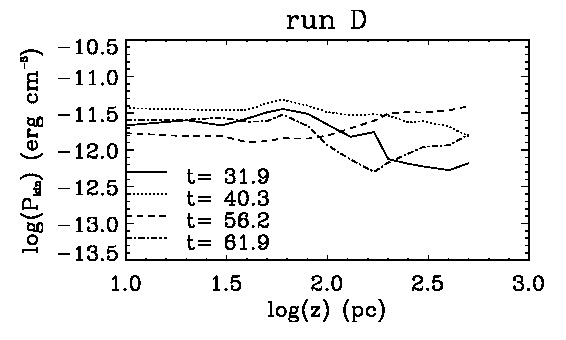}} 
\put(0,13.5){\includegraphics[width=8cm]{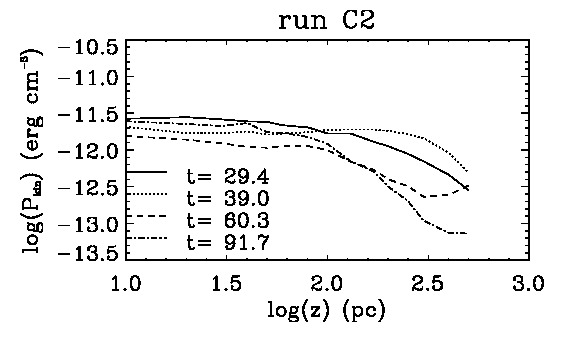}} 
\end{picture}
\caption{Kinetic pressure profile along the $z$-axis for five different models
(see label) at four different timesteps. Run D presents the largest values 
while Run NF1 presents values significantly smaller than the other runs.
Moreover kinetic pressure quickly drops at higher altitude for run NF1.}
\label{Pkin_z}
\end{figure}

\setlength{\unitlength}{1cm}
\begin{figure} 
\begin{picture} (0,18.5)
\put(0,13.5){\includegraphics[width=8cm]{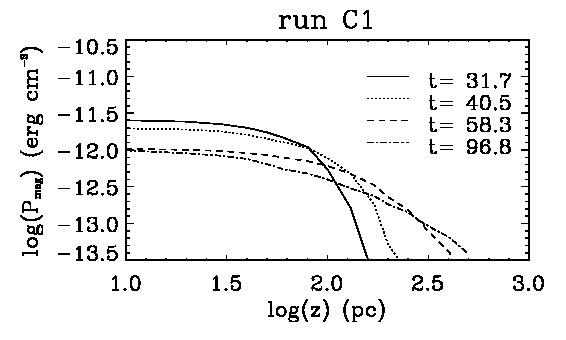}} 
\put(0,0){\includegraphics[width=8cm]{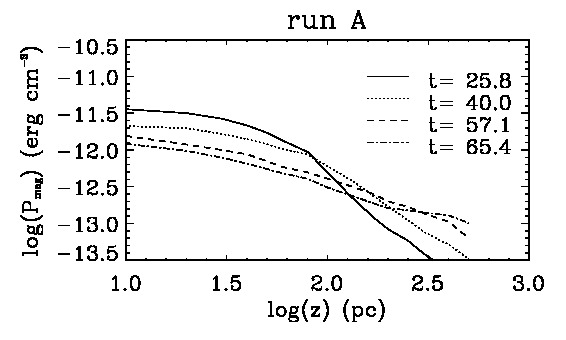}} 
\put(0,4.5){\includegraphics[width=8cm]{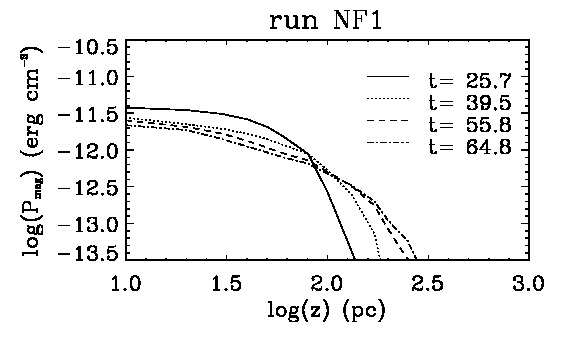}} 
\put(0,9){\includegraphics[width=8cm]{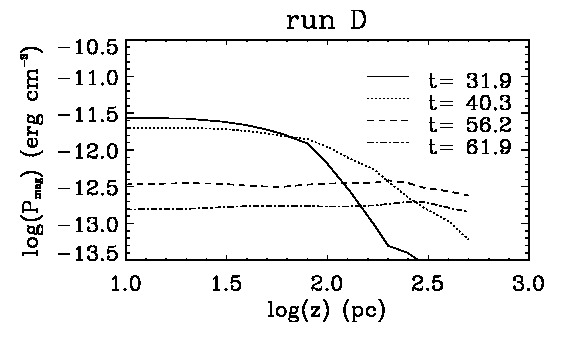}} 
\end{picture}
\caption{Magnetic pressure profile along the $z$-axis for four different models
(see label) at four different timesteps. All runs show similar values
in the midplane. For runs A, C1 and D the profile tends to become 
shallower with time illustrating the transport of the magnetic flux
toward higher altitude.}
\label{Pmag_z}
\end{figure}

We show four timesteps for each runs. The first one is at about 
25-30 Myr, the second at 40 Myr and the third at about 55-60 Myr. For the fourth
we select a timestep of about 90 Myr for runs C1 and C2 and about 65-70 Myr for 
run A, D and NF1. The first timesteps have been chosen at the beginning 
of the star forming phase (see Fig.~\ref{mass_sink_tot}) and the second at about 10 Myr
later because the SFR is typically close to its maximum.
 For the third timesteps  the SFR is nearly constant with time  and thus the profiles
 correspond to the quasi-stationnary regime. This is confirmed by 
the last  timestes which show no significant evolution with respect to the third one
although for runs A, D and NF1 the evolution is faster because accretion is higher
and the profiles keep evolving rapidly at later times.

For the sake of conciseness, we will select  the runs which we
think are most relevant to emphasize the impact of the physics and of the schemes.
When investigating the global disk structure (sect.~\ref{global}), we concentrate mainly 
on runs NF1, A, C1, C2 and D. When we investigate the multiphase ISM (sect.~\ref{multiphase}), 
we restrict to runs C1 and C2. The other runs (NF2, C3 and C4) are used to 
quantify the influence of the supernovae scheme on the star formation rate (sect.~\ref{sfr}).

\section{Global structure}
\label{global}
In this part we characterize the global structure of the galactic disks.

\subsection{Qualitative description}

Figure~\ref{image} shows the  column density field along the 
$z$-axis (left panels) and $y$-axis (right panels) for 
MHD run C1 (upper panels) and hydrodynamical run C2 (lower panels).
In both cases, the disk is clearly visible although its structure
is quite irregular and varies significantly from place to place. 
The disk is slightly thicker in the magnetized run than in the 
hydrodynamical one, which is a clear consequence of the magnetic support. 
The  column density distributions appear
different with and without magnetic field. In particular, 
 small scale  fluctuations are more pronounced in the 
hydrodynamical run. This trend is also similar to what has been 
found at smaller scales \citep[e.g.][]{h+2008} and has been interpreted 
by \citet{h2013} to be  a consequence of the Kelvin-Helmholtz 
instability being stabilized by the  magnetic field \citep{ryu+2000}. As will be discussed 
later in the paper, this has consequences on the mass of self-gravitating 
objects, which form. 

Figure~\ref{image2} shows column densities for the run without 
supernovae feedback, run NF1 (upper panels), and with our feedback scheme D
as described above. The disk structure is very different 
in both cases.  When no feedback is applied, very long filaments
develop across the computational box. They converge towards a
region where most of the mass accumulates. There are much less 
dense regions compared to  run C1 and the disk is obviously thinner. 
This behaviour is very similar to what is reported in \citet{hopkins+2011} (see for 
example their Fig.~2)
On the other hand, scheme D leads to a disk whose structure is much more irregular 
than the structure of the disk obtained with scheme C although the 
number of supernovae and their energy are identical in the two cases. 
This constitutes a clear confirmation to the works of previous authors \citep[e.g.][]{bournaud+2010, dobbs+2011, hopkins+2011}
that feedback is playing a crucial role for the structure of galactic disks and 
(see next section) for regulating  star formation. It is also clear 
that the correlation between the gas and the supernovae does influence 
significantly the galactic disk structure.

\subsection{Disk density profile}
One fundamental aspect for the galactic structure is the density 
profile and the typical thickness of the gas distribution.
In the case of the Milky Way, it has been measured (e.g. Ferri\`ere 2001)
that the different phases have different distributions. They tend to be
roughly Gaussian but their thickness vary.  
 The molecular  gas has a full width at half maximum (FWHM) of about 
120 pc while for the atomic gas  it is about 230 pc. Both the 
molecular and the atomic gas have a 
mean density of about 0.5 cm$^{-3}$ for a total of 1 cm$^{-3}$.

Figure~\ref{rho_z} shows the density distribution along the $z$-axis 
for five different models. The run C1 presents a density of about 3-4 cm$^{-3}$
and a FWHM of about 50 pc. 
This is also roughly the case for the hydrodynamical run. 
Comparison with run C1b shown in the appendix
reveals that numerical resolution may be partly at the origin of this discrepency.
In the run based on scheme D, the maximum density varies with time because 
of stronger accretion but is around 1-2 cm$^{-3}$ after 40 Myr. The FWHM
is  larger than in the previous cases and equal to about 100 pc. 
When no feedback is included, the disk tends to be  thinner. 
For example at times 55.8 and 64.8 Myr, it is about 30 pc. 
A similar distribution is found at time 57.1 and 65.4  Myr with scheme A, 
which as we will see later has an accretion behaviour very similar 
to the run without feedback.

\subsection{Pressure support}
Since the disk thickness is a direct consequence of the 
various supports, we present the profile along $z$-axis 
of the three relevant 
pressures, namely  thermal,  kinetic and  magnetic
ones noted $P_{th}$, $P_{kin}$, $P_{mag}$ respectively. 
While  $P_{th}$ and $P_{mag}$ have their 
standard definitions, $P_{kin}$ is taken as 
\begin{eqnarray} 
P_{kin} = { \sum v_z^2 \rho dV \over \sum dV }. 
\end{eqnarray}

Figures~\ref{Pth_z},~\ref{Pkin_z} and~\ref{Pmag_z}
show $P_{th}$, $P_{kin}$, $P_{mag}$ respectively, 
as a function of  altitude. 

In all models, the thermal pressure ranges between a few $10^{-13}$ and 
a few $10^{-12}$ erg cm$^{-3}$, 
which corresponds to the typical pressure in the ISM \citep[e.g.][]{ferriere+2001}.
Its variation  with the altitude,  $z$, closely follows the density variation
with the notable exception of runs NF1 and  A (i.e. no or randomly distributed supernovae). 
For this latter, the 
thermal pressure is roughtly constant up to an altitude at which the density 
is about 0.1 cm$^{-3}$. This is clearly due to the  an efficient 
production of  warm and hot gas by supernovae explosions.
Note that run D presents more variability than run C1. This is likely a consequence 
of the supernovae being less spatially correlated to the sink particles. When a supernova
explodes in a dense regions, it tends to mimic what happens for most supernovae in run C1 but when it explodes in a diffuse regions, it tends to mimic 
run A where most supernovae explode in the WNM which has the largest volume filling factor.

The kinetic pressure is typically a few times larger than the thermal one
and depending of the model, reaches values of the order of $\simeq 1-3 \times 10^{-12}$
erg cm$^{-3}$. These values are also very similar to what is reported in 
related studies \citep[e.g.][]{joung+2009,kim+2011}. Interestingly, the scale height 
of $P_{kin}$ is larger than the density scale height by a factor on the order of 1.5-2 for run C1 and up to 3-4
for run C2.  In the
case of scheme D, it even slightly increases with altitude. 
As expected, in the absence of supernovae feedback, $P_{kin}$ drops 
to small values rapidly.
  
The magnetic pressure is comparable to the thermal pressure
and reaches values of the order of a few  $10^{-12}$ erg cm$^{-3}$.
It therefore contributes  to support the galactic disk
against gravity. Interestingly, while the 
magnetic intensity tends to decrease in the midplane 
as time goes on, it tends to increase with time 
at high altitude. This is a consequence of the generation 
of magnetic field through turbulence but also 
a consequence of the transport of the field lines by galactic outflows. 

\setlength{\unitlength}{1cm}
\begin{figure} 
\begin{picture} (0,9.5)
\put(0,4.5){\includegraphics[width=8cm]{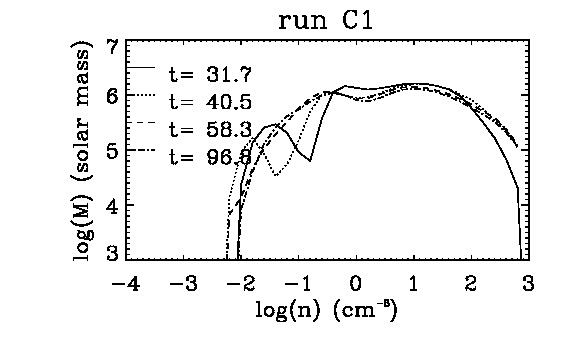}} 
\put(0,0){\includegraphics[width=8cm]{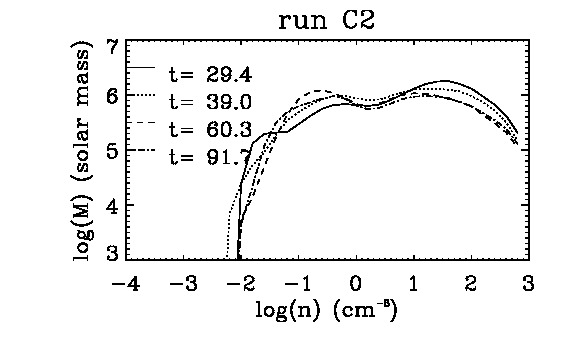}} 
\end{picture}
\caption{Density PDF for MHD run C1  (upper panel) and  hydrodynamical
run C2 (lower panel). The small drop at $n \simeq 2-3$ cm$^{-3}$ corresponds
to the thermally unstable regime, which persists in spite of the strong turbulence. The two peaks
correspond to the WNM and CNM phases.} 
\label{mass_dens}
\end{figure}

\setlength{\unitlength}{1cm}
\begin{figure} 
\begin{picture} (0,9.5)
\put(0,4.5){\includegraphics[width=8cm]{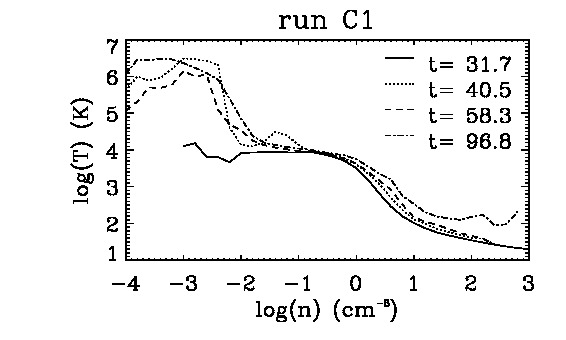}} 
\put(0,0){\includegraphics[width=8cm]{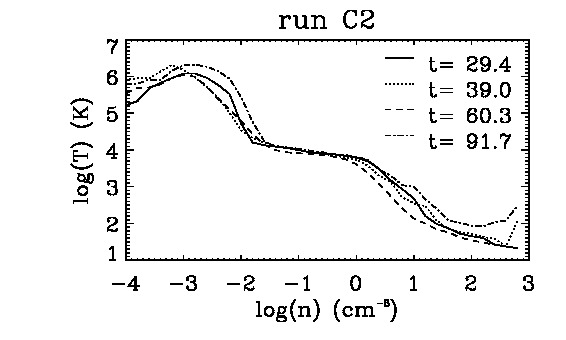}} 
\end{picture}
\caption{Mean temperature as a function of density for run C1 (upper panel) and C2
 (lower panel). As expected the temperature lies mainly in 3
ranges of temperature  namely $10^6$, $10^4$
and $10^2$ K, corresponding to the three phases of the ISM, the HIM, the WNM and the CNM.}
\label{T_dens}
\end{figure}

\setlength{\unitlength}{1cm}
\begin{figure*} 
\begin{picture} (0,18)
\put(0,8){\includegraphics[width=9cm]{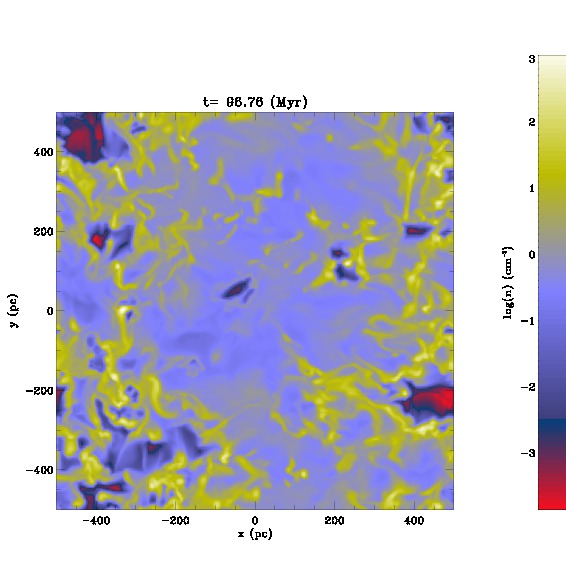}} 
\put(9,8){\includegraphics[width=9cm]{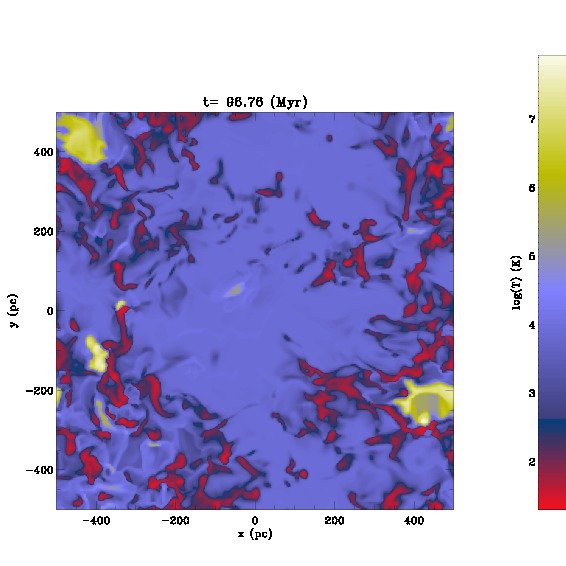}} 
\put(0,0){\includegraphics[width=9cm]{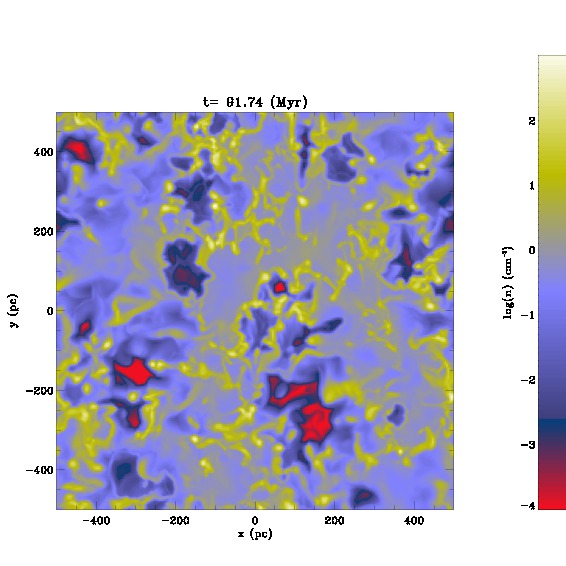}}  
\put(9,0){\includegraphics[width=9cm]{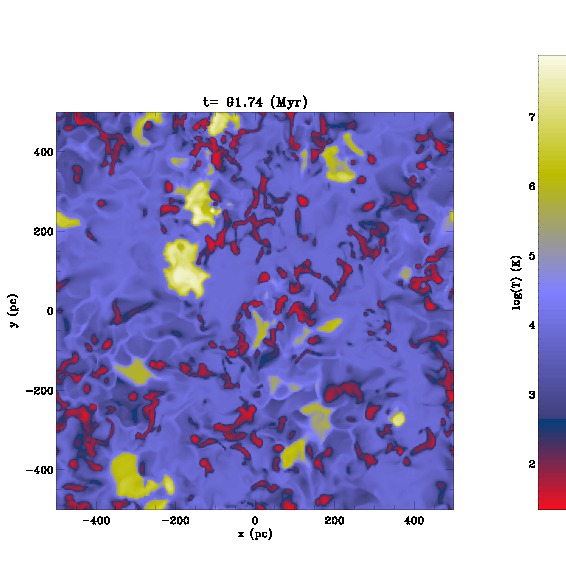}}  
\end{picture}
\caption{Density (left column) and temperature (right column) fields
in the equatorial plane for the MHD run C1 (upper pannel)
and the hydrodynamical run C2 (lower panels). The figures illustrate the 
multi-phase nature of the interstellar medium. Most of the volume is filled 
by warm neutral gas with temperature of about 8000 K. A tiny fraction is occupied by 
the hot phase at temperature larger than $10^6$ K. }
\label{im_dens_T}
\end{figure*}

\setlength{\unitlength}{1cm}
\begin{figure} 
\begin{picture} (0,9.5)
\put(0,4.5){\includegraphics[width=8cm]{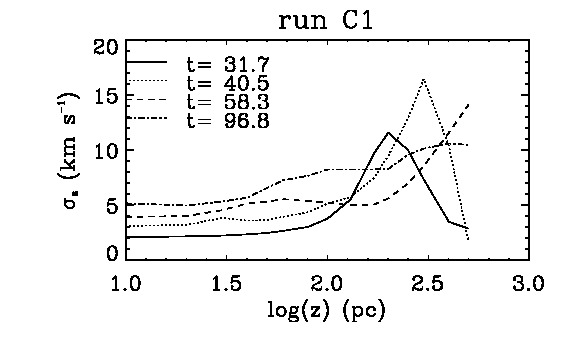}} 
\put(0,0){\includegraphics[width=8cm]{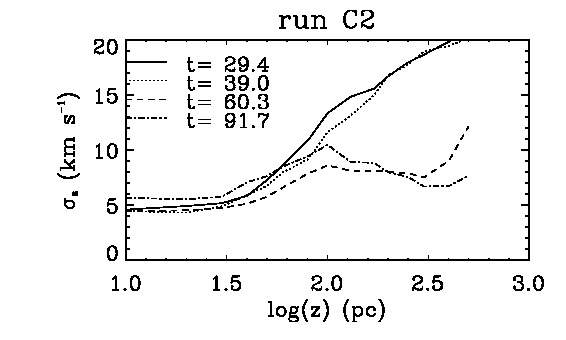}} 
\end{picture}
\caption{Root mean square value of $v_z$ as a function of $z$ (see text)
for MHD run C1  and hydrodynamical run C2. Typical values are 
about 4-5 km s$^{-1}$ at the midplane.}
\label{rmsVz_z}
\end{figure}

\setlength{\unitlength}{1cm}
\begin{figure} 
\begin{picture} (0,9.5)
\put(0,4.5){\includegraphics[width=8cm]{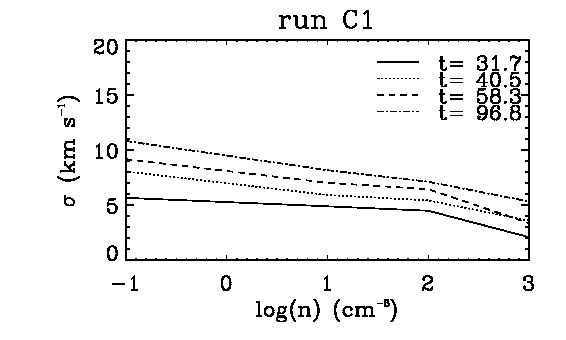}} 
\put(0,0){\includegraphics[width=8cm]{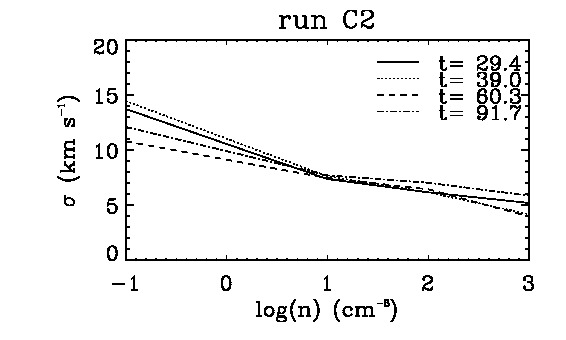}} 
\end{picture}
\caption{Root mean square velocity as a function of density 
for MHD run C1 and hydrodynamical run C2. }
\label{Vrms_dens}
\end{figure}

\setlength{\unitlength}{1cm}
\begin{figure} 
\begin{picture} (0,9.5)
\put(0,4.5){\includegraphics[width=8cm]{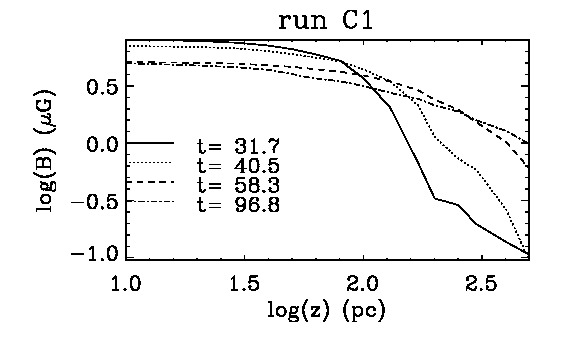}} 
\put(0,0){\includegraphics[width=8cm]{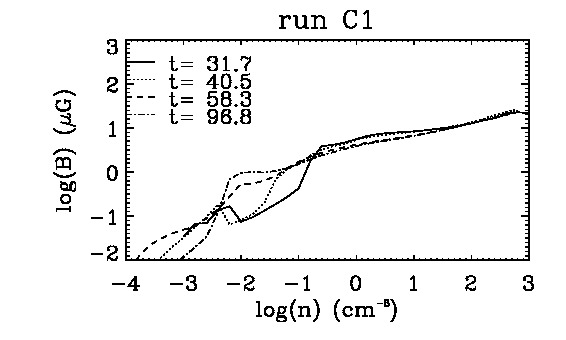}} 
\end{picture}
\caption{Mean magnetic intensity as a function of $z$
and as a function of density. Note in particular that typical values of about 5 $\mu$G are 
being obtained in the midplane. Between 1 and $10^3$ cm$^{-3}$, the magnetic intensity
weakly varies with the density.}
\label{mag}
\end{figure}

\section{Multiphase ISM}
\label{multiphase}
In this section, we study the density, temperature and magnetic field 
distribution of the gas in the simulations. For conciseness, we restrict our attention 
to runs C1 and C2 only stressing the most interesting differences with other runs. 
The comparison between runs C1 and C2  emphasizes the role that the magnetic field has to determine 
the characteristic of the ISM.

\subsection{Density and temperature distributions}
 As it is the case in other works \citep[e.g.][]{deavillez+2005,joung+2006,hill+2012}, 
the interstellar medium which is initially uniform in density and temperature, 
quickly breaks up into a multiphase medium in which the density 
varies from less than 10$^{-3}$ cm$^{-3}$ to almost 10$^3$ cm$^{-3}$
while the temperature can be as high as 10$^7$ K and as low as a few tens 
of Kelvin. 

Figure~\ref{mass_dens}  shows the mass contribution  of the various gas densities in 
the computational box. It is dominated by the dense gas ($n > 10$ cm$^{-3}$) but 
a non-negligible fraction lays at lower densities, which corresponds to the 
WNM regime and thermally unstable gas ($0.1 < n < 10$ cm$^{-3}$). 
Interestingly, there is a small dip in the thermally unstable regime ($n \simeq 2-3$ cm$^{-3}$).
This indicates that in spite of the relatively high level of turbulence, typical of galactic 
disks (see below), the 2-phase  structure \citep{wolfire+2003} that would be obtained in a static medium, is not 
erased  by the dynamical processes and able to persist. At early time there is 
a little more dense gas in the hydrodynamical case. This is due to the magnetic support 
that reduces the amount of self-gravitating gas.  For run D (not displayed here for conciseness), the transition 
between WNM and cold neutral medium (CNM) is less pronounced, which is a consequence of the stronger turbulence in this run.
These behaviours are reminiscent 
of what has been found in colliding flow calculations, which attempted to model the ISM 
at scales on the order of 10-50 pc \citep[e.g.][]{ah05,vazquez+2006,h+2008,heitsch+2008,
banerjee+2009,ah+2010,inoue+2012}
and also in similar supernovae regulated galaxy simulations \citep[e.g.][]{dib+2006,hill+2012,kim+2013}.
In particular, in these simulations it has been found that the ISM quickly breaks up 
into a multi-phase, clumpy medium where the 2-phase behaviour (i.e.  present an excess of gas in thermodynamical states close
to the two stable branches of thermal equilibrium) is maintained even though the 
medium is largely turbulent.

Figure~\ref{T_dens}  displays the temperature distribution  as a function of density.
It clearly shows the existence of 3 main domains corresponding to 
the hot ionised medium ($T \simeq 10^6$ K), the warm neutral medium ($T \simeq 10^4$ K),
and the cold neutral medium ($T<100$ K). This is in good agreement 
with the classical 3 phases model of the interstellar medium as early described for example 
in \citet{mckee-ostriker1977}. 
As noted by previous 
authors \citep[e.g.][]{gazol+2001}, there is gas in the thermally unstable regions 
($T\simeq 10^3 K$), whose existence is permitted by the turbulent motions. 
The two runs present similar distributions although the transition between the warm 
and the cold phase (at density 1-10 cm$^{-3}$) is a little more shallow for the 
MHD run (C1) than for the hydrodynamical one (run C2). This is because the magnetic field 
contributes to the total pressure and can therefore stabilize the pieces of gas that
are thermally unstable.  Again in run D, the 3 regimes  
are less clearly separated though the global temperature range in the simulation is similar.

The spatial distribution is illustrated by 
Fig.~\ref{im_dens_T}, which displays a cut through the equatorial plane 
of the density and temperature fields both for MHD run C1 and hydrodynamical run C2. 
As can be seen, most of the volume is found to be occupied by the warm neutral gas 
with temperature of the order of 10$^4$ K and densities of the order of 
1 cm$^{-3}$.  The hot gas, produced within supernovae explosions, occupies
only a small fraction of the volume. Interestingly, the structures in the 
hydrodynamical and MHD runs have a slightly different shape. 
As already discussed in the previous section, the dense clouds in 
the hydrodynamical simulations tend to be more fragmented and 
on average slightly smaller (see also Fig.~\ref{image}).

\subsection{Velocity dispersion}

In the supernovae regulated numerical simulations \citep[e.g.][]{slyz+2005,deavillez+2005,joung+2006,hill+2012, kim+2011, kim+2013, gent+2013} the kinetic energy, which  decays through the turbulent cascade, is  
replenished by the supernovae explosions.  The velocity dispersion in the 
computational box  is the result of a balance between  injection and dissipation 
as emphasized for example in \citet{maclow+2004}, who present orders 
of magnitude suggesting that supernovae explosions can explain the 
velocity dispersion observed  in the Milky Way to be of the order of 6 km s$^{-1}$. 

Figure~\ref{rmsVz_z} displays the rms $z$-component of the velocity field 
weighted by density 
\begin{eqnarray}
\sigma_z = \sqrt{ { \sum v_z^2 \rho dz  \over  \sum  \rho dz } }.
\label{eq_rmsVz_z}
\end{eqnarray}

Close to the equatorial plane,  $\sigma _z$ is of the order of 4-5 km s$^{-1}$
for  run C1 and 5-6 km $^{-1}$ for run C2.  At 
higher altitude, the velocity dispersion increases to values 
of about 8-10 km s$^{-1}$ at 100 pc for  run C1 and 8-15 km s$^{-1}$
for run C2. This is essentially due to the gas density getting 
lower at higher altitude. 
These values are again similar to what has been previously reported in similar studies
\citep[see e.g. Fig.~4 of] []{kim+2013} and can be understood by relatively simple
considerations.  Following \citet{maclow+2004}, we can simply estimate the amount of mechanical energy which dissipates
in the turbulent cascade as $\dot{E}_{diss} = M \sigma^2 / \tau$ where 
$M$ is the total mass of the system, $\tau$ is the crossing time and $\sigma$ the total velocity 
dispersion. Assuming that 
$\tau = h / \sigma$, where $h$ is the disk scale height, we get  $\dot{E}_{diss} = M \sigma^3 / h$
where $\dot{E}_{diss}$ is the energy dissipated per units of time.
The amount of energy which is injected by the supernovae into the system is simply 
$\dot{E}_{inj} =  \epsilon \dot{N}_{sn} \times 10^{51} \, \rm{erg}$, where 
$\dot{N}_{sn}$ is the density of supernovae per units of time and $\epsilon$ is the efficiency
at which turbulence is triggered. Equating these two rates, we get 
\begin{eqnarray}
\sigma = \left( {\epsilon \dot{N}_{sn} h \times 10^{51} \, \rm{erg} \over M } \right)^{1/3}.
\end{eqnarray}
To estimate the value of $\sigma$, we take values typical for the Milky way. These values are also 
representative of our simulation parameters. 
We take  a mass of $10^{10} M_\odot$, a frequency of supernovae  $\dot{N}_{sn}=1/50$ yr$^{-1}$, 
a height $h=100$ pc and an efficiency  $\epsilon=0.1$, we get  
\begin{eqnarray}
\sigma \simeq 8 \, {\rm km \, s^{-1}} \left( {\epsilon \over 0.1 } {\dot{N}_{sn} \over 1/50 {\rm yr}^{-1} } {h \over 100 {\rm pc}} 
{10^{10} M _\odot \over M } \right)^{1/3}.
\end{eqnarray} 
This value is thus in good agreement with the velocity dispersion inferred from the simulations and from the observations.
It is worth stressing that due to the weak dependence in all parameters (to the power 1/3), it is relatively unsurprising 
to find that the velocity dispersion generally does not undergo large variations. 
We note that the fluctuations in the hydrodynamical 
run C2 (lower panel) appear to be quite large with respect to the MHD run C1 (upper panel). 
This is likely a consequence of the higher SFR (see below) found in the hydronamical case. 
This results in a stronger feedback.  

Finally, since stars mainly form in the dense gas, it is important to understand the star formation process, to know more 
accurately how velocity dispersion depends on the gas density. For that purpose 
Fig.~\ref{Vrms_dens} displays the rms velocity field 
(taking into account the 3 components) weighted by density as a function of  density
in the whole computational box. As expected the velocity dispersion 
is weaker in the dense gas than in the diffuse one by a factor of about 2.  Typical velocity 
dispersion in the dense gas is  on the order of 4-5 km s$^{-1}$.

\subsection{Distribution of magnetic intensity}

Figure~\ref{mag} shows the magnetic intensity as a function of altitude (upper panel)
and density (lower panel).  At the mean density of the galactic disk, i.e. $n \simeq 2-3$ cm$^{-3}$
the magnetic intensity is about 3-5 $\mu$G, which is also the mean value  up to an  altitude of about 100 pc. 
This is coherent with the values of about 5 $\mu$G reported by \citet{heiles+2005}.
At higher densities, the magnetic intensity increases and reaches values of about 10$\mu$G 
at densities of about $10^2$ cm$^{-3}$. Note that this  corresponds to a rather shallow variation 
of the magnetic intensity with density as observed in the diffuse gas \citep[e.g.][]{troland+1986}.
The exact reason of this weak correlation is most likely due, on the one hand to the Lorentz  force, 
that resists contraction perpendicular to the field lines \citep{h+2000,passot+2003}. On the other hand, it is 
also partly due to the turbulent diffusivity, which has also been observed to play an 
important role in numerical simulations \citep[e.g.][] {lazarian+1999,santos-lima+2010,h+2011}.
 For densities below $n \simeq 10^{-2}$ cm$^{-3}$, a steep drop is observed with density. This is 
due to the fast expansions produced by supernovae explosions, which tend to dilute the magnetic intensity
 very significantly.

\section{Star formation rate, sink mass function and outflows}
\label{sfr}
We now investigate the characteristics of the star formation in the simulations.
This is achieved through the sink particles described in sect.~\ref{code}. 
We first quantify the total mass of the sink particles, which represents the 
star formation rate in the simulations. We then study the mass distribution, i.e.
the sink mass function of some of our models. Finally, we study the outflows
which are launched at high altitude and eventually escape the computational box.

\subsection{Star formation rates}

\setlength{\unitlength}{1cm}
\begin{figure} 
\begin{picture} (0,15)
\put(0,10){\includegraphics[width=9cm]{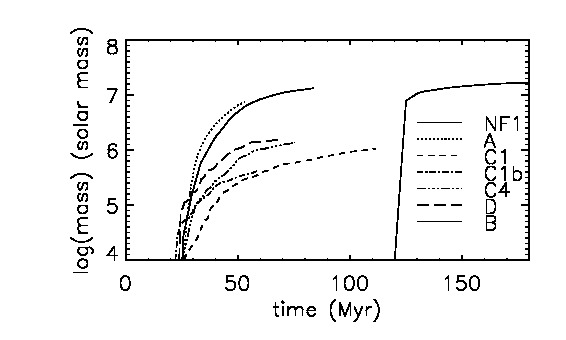}} 
\put(0,5){\includegraphics[width=9cm]{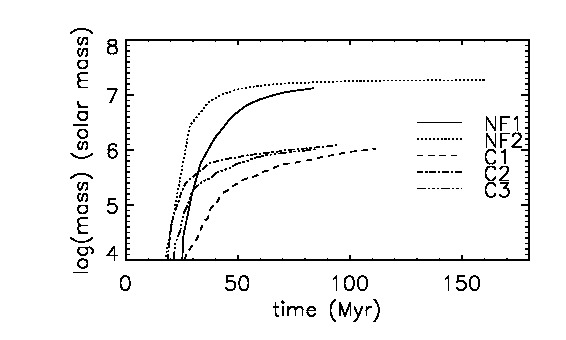}} 
\put(0,0){\includegraphics[width=9cm]{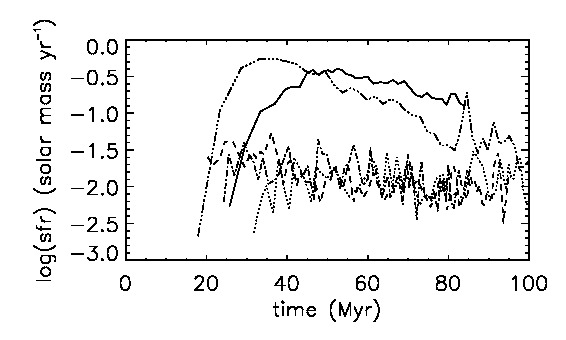}} 
\end{picture}
\caption{Total mass of sink particles as a function of time for the various models.
Upper panel shows the influence of the supernova feedback scheme while 
middle and lower panels show the influence of the magnetic field. Lower panel 
shows the SFR (i.e. time derivative of mass) corresponding to middle panel.
In upper panel, run B corresponds to the solid curve which starts at 120 Myr.
The runs with feedback present SFR that are typically 10 to 30 times 
smaller. Magnetic field reduces the SFR by a factor on the order of 2.}
\label{mass_sink_tot}
\end{figure}

Figure~\ref{mass_sink_tot} shows the total mass of the sink particles as a function of time 
in the simulations. Upper  panel shows the influence of the sink particle prescription
while lower panel shows the influence of the magnetisation. 
Before commenting on the difference between the various models, we first discuss the main trends
and numbers. For all models (except run B) accretion onto sink particles starts between 
 20 and 30 Myr. In about 10-20 Myr, the total accreted mass varies between
a few 10$^5$ to  $\simeq$10$^7$ $M_\odot$. As discussed below, these differences are due to 
the various feedback prescriptions and magnetisations. At later times, all models tend to reach 
a phase of stationary accretion at a rate which ranges from about 10$^{-2}$ to $\simeq 10^{-1}$ $M_\odot$ yr$^{-1}$. 
It is worth comparing these values with the typical 3 $M_\odot$ yr$^{-1}$ at which  Milky way is forming stars. 
In order to do so, one must first correct for the volume of our computational box. Since in the Milky way
most stars form inside the solar circle whose radius is about 8 kpc and 
since the size of the computational domain is equal to 1 kpc, 
a geometrical factor of $\pi \times 8^2 \simeq 200$ should be taken into account.
However, it should also be accounted for the fact that the efficiency of the mass eventually accreted into 
the stars is only a fraction of the mass that is accreted onto the cores. This value 
is not known with great accuracy but has been estimated to be of the order of 
1/3 \citep[e.g.][] {alves+2007}. We note that the sink particles used in this study are at this stage
much larger than dense molecular cores. Therefore it could be that the efficiency should be even 
lower than this value. Combining these two numbers, we find that for a galaxy like the Milky way, 
our models would predict a star formation rate ranging from about 1 to 20 $M_\odot$ yr$^{-1}$.
Given the large uncertainties, the first value appears to be in reasonable agreement with the 
galactic one.

\subsubsection{Influence of feedback prescriptions}
All models displayed in top panel of Fig.~\ref{mass_sink_tot} have initial conditions identical to 
 run C1, i.e. have an initial magnetic field 
whose initial intensity in the midplane is about 2.5$\mu$G.

First of all, the large difference between the solid line (run NF1) and 
the dashed line (run C1) confirms the drastic influence of the feedback on 
the star formation rate which is reduced by a factor of 20-30. This constitutes 
a strong hint that feedback can be largely responsible to solve 
the long standing issue of the so-called Zuckermann-Evans 
catastroph \citep{zuckerman+1974}. If all the molecular gas of the Milky way was 
collapsing in a free-fall time, about 300  $M_\odot$ yr$^{-1}$ of stars 
would form in the Galaxy. 

Second of all, when the supernovae are not correlated to density  (run A)
not only is the feedback  unable to reduce the star formation rate but this latter 
is even slightly higher.  This is because since most of the volume is occupied 
by warm gas, most of the supernovae therefore explode in  low density regions.
Their net effect is thus to further compress the dense gas. When the supernovae 
are correlated with the density peak  (run B), it takes a long time 
before sinks can form because the dense gas is efficiently dispersed. However, 
once sinks start forming, since the supernovae are not correlated locally 
in space and in time with accretion 
but simply with the densest cell in the simulation, they are unable to reduce the 
accretion rate. Therefore  star formation rates comparable to the run
without feedback are obtained. 

Third, the runs C4
and  D  show that  SFR larger 
by a factor 2-3 are obtained when the feedback is either purely thermal 
or less tightly correlated to the sink particles. Given that these 
two aspects are largely uncertain, this illustrates the 
limit of this modeling and suggests that the typical accuracy of these models 
is at best on the order of a factor 2-3. Note that another severe source of 
uncertainties comes from the time at which supernovae are introduced. In particular, 
if a delay of tens of Myr is introduced,  SFR comparable to the ones of run NF1 are obtained.
This  suggests that in order to get 
more accurate models, it is necessary to have a better description of the small 
scales and in particular of the formation and evolution 
of massive stars up to the point where they explode. Ideally, this would require 
running a set of specific small scale simulations to quantify more accurately the 
impact of the feedback.

\subsubsection{Influence of magnetisation}
All models displayed in middle and bottom panel  of Fig.~\ref{mass_sink_tot} are performed 
either with no feedback (runs NF1 and NF2) or with the 
same feedback scheme (scheme C).  Different levels of magnetisation are compared. 

In the hydrodynamical 
run C2, stars start forming a few Myr before 
 run C1. The SFR is initially 
significantly reduced compared to run C2. At later time, 
they become however comparable. These effects are  a consequence
of the magnetic support which contribute to resist the gravitational   
contraction but also to the  density PDF that is narrower in the presence 
of a magnetic field \citep{molina+2012}. A similar effect
 is obtained for 
the two runs without feedback (runs NF1 and NF2) for which it is seen 
that the SFR is a little higher in the hydrodynamical case than in the MHD one.

Interestingly, even when the magnetic field is rather weak (0.5 $\mu$G initially), 
it still has a visible impact and reduces the SFR by a factor of about $50 \%$
during the first 20 Myr after star formation has started. This is because
magnetic field is quickly amplified to larger values.

This effect is quantitatively comparable to what has been inferred at smaller 
scales by various teams who investigate star formation in substancially magnetized,
though supercritical clouds. For example \citet{price+2008} 
simulate the collapse of a self-gravitating clump while \citet{dib+2010} 
and \citet{padoan+2011} perform self-gravitating, MHD calculations 
within periodic boxes. They all infer that magnetic field reduces the 
SFR by a factor of about 2. The exact reason of this lower 
value has not been analysed in great details so far but it is likely a consequence 
of the magnetic support which tends to resist gravity and the somehow narrower density PDF
which tends to reduce the SFR \citep{hc2013}.

\subsection{Mass function of sink particles}

\setlength{\unitlength}{1cm}
\begin{figure} 
\begin{picture} (0,20)
\put(0,15){\includegraphics[width=9cm]{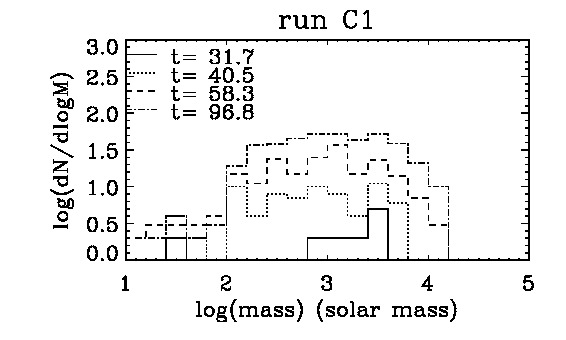}} 
\put(0,10){\includegraphics[width=9cm]{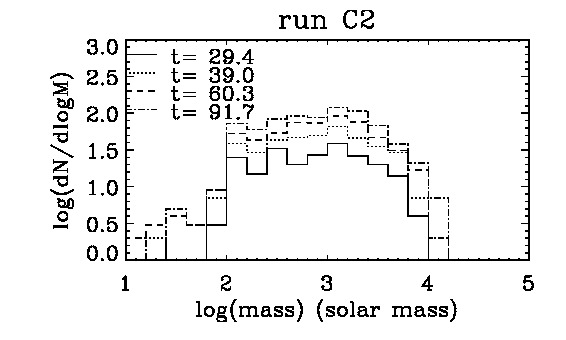}} 
\put(0,5){\includegraphics[width=9cm]{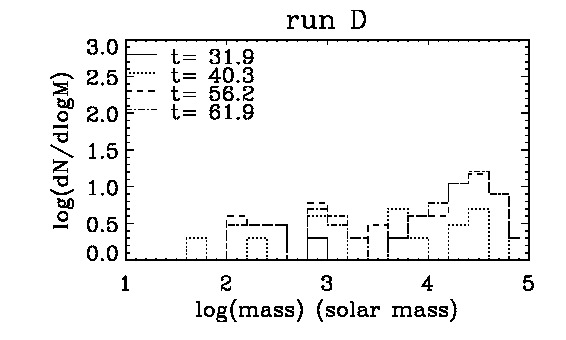}} 
\put(0,0){\includegraphics[width=9cm]{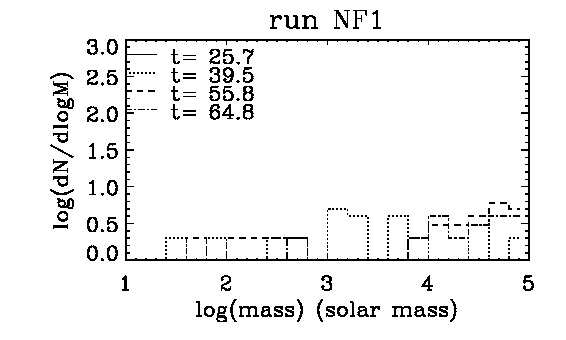}} 
\end{picture}
\caption{Sink mass spectra for four models at four timesteps. 
In the case with no feedback there are typically few massive sink particles while a broader
distribution develops when feedback is included.} 
\label{sink_mass_spectrum}
\end{figure}

Figure~\ref{sink_mass_spectrum} shows the sink mass function for various models
and at four timesteps from which one can verify that the trends discussed below are not due to a time selection. 
For  runs C1 and C2, 
a large number of sinks form (about 400 and 700 respectively for run C1 and run C2).
Their masses span about 3 orders of magnitude. Given the limited numerical resolution 
of the present study many features of the distribution must be taken with great care.
In particular the peak at about 10$^3 \, M_\odot$ would certainly shift to smaller values
in  more resolved runs \citep[e.g.][]{ha07}.
There is a possible 
trend for a powerlaw developing at large masses (in the range $\simeq 3 \times 10^3-10^4 \, M_\odot$)
with an exponent compatible with $\simeq -1$. However, the limited resolution precludes a 
firm conclusion. We note that in massive collapsing magnetised clumps, the number of fragments
has also been found to be reduced by a factor of about two. This is a clear consequence
of the cold gas being more coherent and less fragmented. As noted previously, the 
reason is that the magnetic field makes the flow more coherent since it tends to 
connect fluid particles linked by the magnetic field lines 
 \citep[e.g.][]{h2013}.

The sink mass function obtained when no feedback (run NF1) is included is quite different. There are much less 
sink particles (about 70) and most of 
them have a mass larger than $10^4 \, M_\odot$. Indeed, the most massive 
sink particle has in this case a mass equal to a few 10$^6 M_\odot$. This behaviour 
is again a consequence of the absence of feedback. The gas tends to concentrate 
in a few locations under the influence of gravity.
The sink mass function obtained for run D is inbetween the one obtained for run C1 and run NF1. 
This illustrates again the fact that in  run D, the stellar feedback is less efficient in 
supporting the gas against gravitational collapse.

\setlength{\unitlength}{1cm}
\begin{figure} 
\begin{picture} (0,18.5)
\put(0,13.5){\includegraphics[width=8cm]{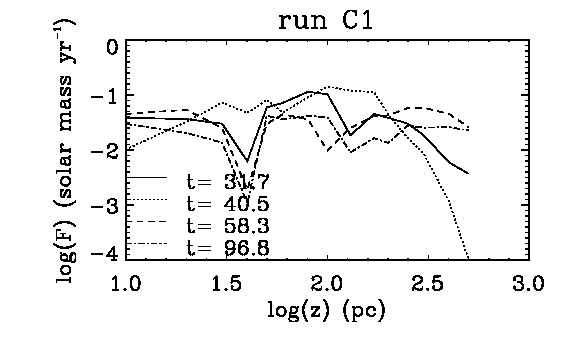}} 
\put(0,0){\includegraphics[width=8cm]{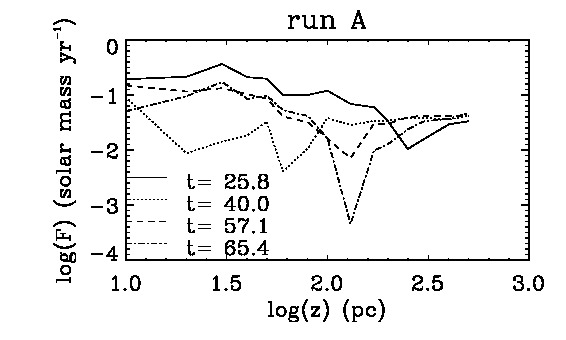}} 
\put(0,4.5){\includegraphics[width=8cm]{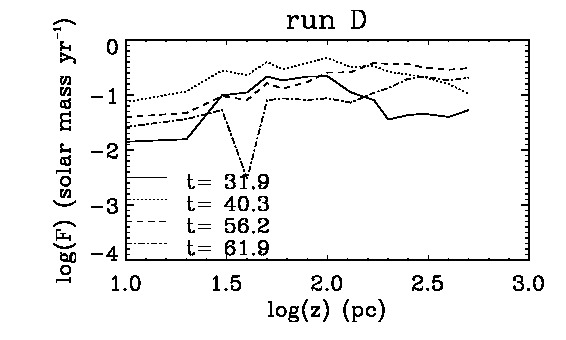}} 
\put(0,9){\includegraphics[width=8cm]{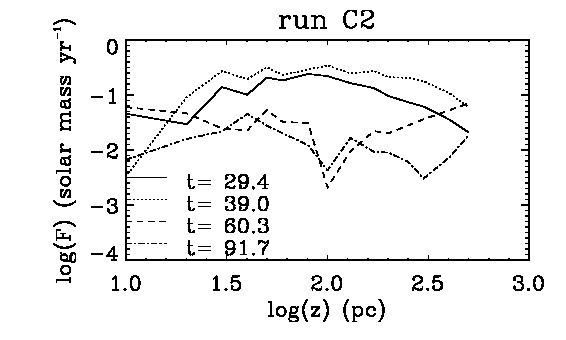}} 
\end{picture}
\caption{Mean flux of mass along the $z$-axis for five different models
(see label) at four different timesteps. The largest values are obtained for 
scheme D.}
\label{flux_z}
\end{figure}

\subsection{Galactic outflows}

The existence of galactic outflows is now well established 
\citep[see][for a recent review]{veilleux+2005} in many galaxies. 
The typical scale height at which these outflows
are observed, is of the order of 
several tens of kpc, which is much larger than the scale 
of the present simulation. Also the box length is equal to only 
1 kpc and we do not have a proper halo structure which 
influences the flow launching \citep[e.g.][]{dubois+2008}. It is nevertheless worth to 
quantify them, in particular because supernovae are believed 
to be largely responsible of their launching.  

Figure~\ref{flux_z} shows the mass flux as a function of the 
altitude $z$ for four timesteps and four models. As can be seen 
it varies significantly with time and altitude from typically a few  
$10^{-2}$ to a few $10^{-1}$~M$_\odot$~yr$^{-1}$. Taking 
into account that the surface of the box is 1 kpc$^2$, this would 
lead to a flux of about $\pi \times 8^2 \simeq 200$ times larger for a galaxy 
similar to ours, i.e. a few solar mass to a few tens or solar mass per year. 
These values are typical of what is measured for galactic outflows
\citep{veilleux+2005}. 

Another interesting trend is that the mass flux broadly correlates
with the SFR (see fig.~\ref{mass_sink_tot}). For 
run C1 (top panel), the peak value of the mass flux is 
about 3 times smaller than what is obtained for run 
D (third panel). Comparing the mean value at the
computational box edges ($z$=500 pc), the ratio 
between the fluxes of run C1 and D leads to somewhat larger values
of about 5-10. 
This is likely a consequence of the dual role of supernovae 
explosions, which are responsible for the regulation of 
star formation through energy and momentum injection in the dense gas 
but also for the launching of the galactic outflows through 
injection  onto the diffuse gas. Since the two processes are linked, 
it is expected that larger SFR lead to stronger outflows as
they imply more feedback. 
Indeed 
the SFR ratio for run D and C1 is about 3 (from upper panel of Fig.~\ref{mass_sink_tot}), 
which is comparable with the value of 3 quoted above but a little to 
small to explain the second value obtained at the box edges. 
This may indicate that another effect must be considered. 
We believe that since  in run D the supernovae
explode further from the sink particles than in run C1,
 more energy and momentum tend to be injected in the diffuse gas than
in run C1. Since the outflows are primarily made by diffuse fastly 
expanding material,  it seems reasonable that the efficiency 
with which they are produced is higher in run C1 than in run D.

\section{Conclusion}

We have performed a series of numerical simulations describing a galactic disk regulated by supernovae feedback at kpc scale. 
Our simulations include both magnetic field and self-gravity. In particular we have explored the influence of various schemes 
to prescribe the supernovae feedback. Our simulations reproduce many features already found by other authors such as 
multi-phase density and 
temperature distributions or velocity dispersion, typically of the order of 5 km s$^{-1}$ in the galactic plane.
Our results are as follows. When the supernovae are randomly distributed they drive the interstellar turbulence but are unable 
to resist self-gravity efficiently and the star formation rate is as high (even slightly higher) as  when no feedback is included. 
When supernovae are correlated to the density peaks, they efficiently limit star formation by preventing the gas to become too dense. 
However as time goes on, dense gas eventually develops. When sink particles are being introduced, then the star formation rate is 
as high as its value without feedback.

When supernoave are spatially and temporally correlated to star formation events, the star formation rate is significantly 
reduced by a factor of the order of ten or more. However, we find that the exact implementation of the supernovae does influence 
the galactic disk structure and the star formation rate significantly. In particular, if the supernoave are distributed in a shell 
of about 16 pc around the sink particles, the accretion rate is higher by a factor of about 3 than if they are randomly placed 
within a sphere of radius equal to 16 pc. In a similar way, if the feedback is purely thermal, the star formation rate is about 2 
times larger than if it has 5\% of kinetic feedback. This implies that a detailed knowledge of how the feedback 
operates on small scales is mandatory to understand its impact with sufficient precision. In particular, the correlation 
between the massive stars and the dense star forming gas should be determined using small scale simuations.

Magnetic field has a significant impact. It delays and reduces star formation by a factor of the order of 2. 
It also tends to reduce the number of star formation regions (e.g. sink particles) by a factor of about 2
therefore producing slightly bigger star forming regions.  
Finally, it should be kept in mind that magnetic field has an important impact on the fragmentation 
of massive cores that it tends to reduce significantly \citep{commercon+2011,myers+2013}.
This implies that more massive stars form when magnetic field is strong. Since feedback is 
a non-linear function of the stellar masses and since feedback influences drastically 
the galactic structure and evolution, it is likely the case that the impact magnetic field 
has on  galaxy evolution is probably even larger than what is  estimated here.  
\\ \\

\emph{Acknowledgments}
We thank the anonymous referee for a careful reading of the manuscript which has 
significantly improved the paper.
This work was granted access to HPC resources of CINES under the 
allocation  x2014047023 made by GENCI (Grand Equipement National de Calcul Intensif).
PH acknowledge the finantial support of the Agence National pour la Recherche through the 
 COSMIS project.
This research has received funding from the European Research Council under the European
 Community's Seventh Framework Programme (FP7/2007-2013 Grant Agreement no. 306483 and no. 291294).

\section{Appendix: the issue of numerical convergence}

\setlength{\unitlength}{1cm}
\begin{figure} 
\begin{picture} (0,23.)
\put(0,18){\includegraphics[width=8cm]{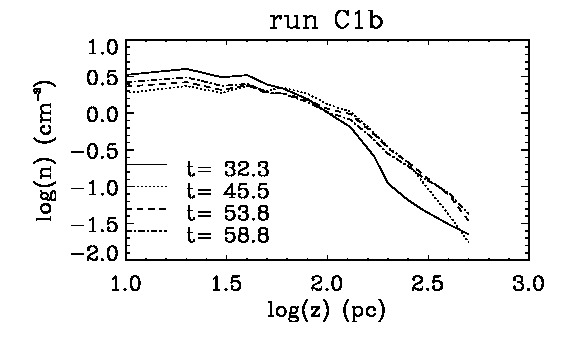}} 
\put(0,4.5){\includegraphics[width=8cm]{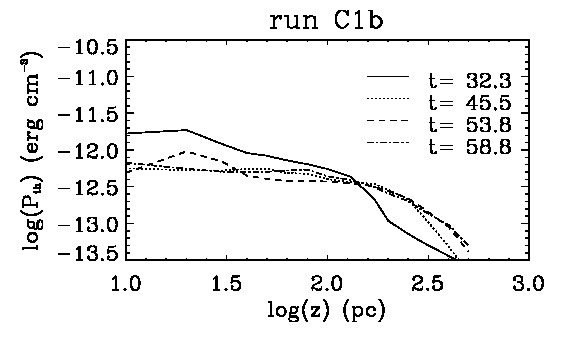}} 
\put(0,9){\includegraphics[width=8cm]{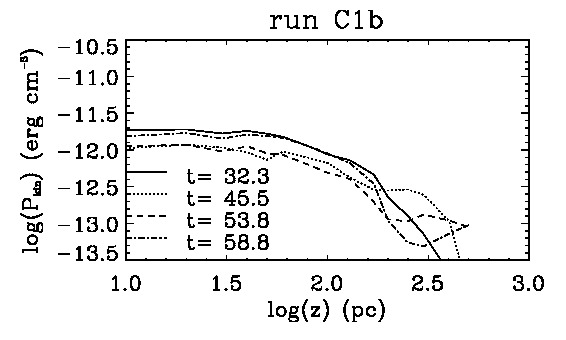}} 
\put(0,0){\includegraphics[width=8cm]{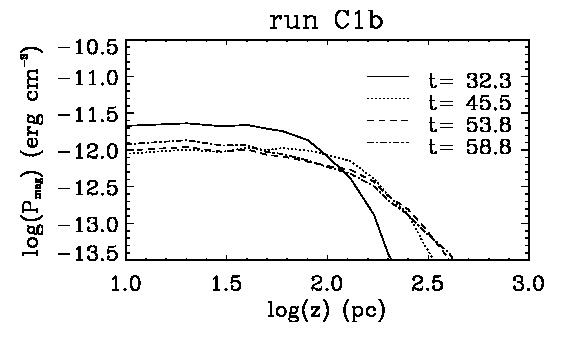}} 
\put(0,13.5){\includegraphics[width=8cm]{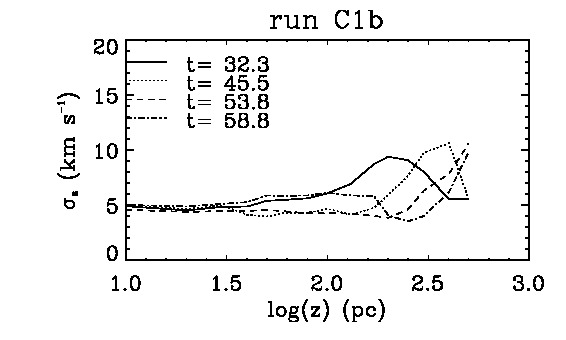}} 
\end{picture}
\caption{Mean density, $z$-velocity dispersion, kinetic, thermal and magnetic profile along the $z$-axis for run C1b
 at four different timesteps. These results should be compared with the corresponding quantities for run C1
displayed in Fig.~\ref{rho_z}, \ref{Pth_z}, \ref{Pkin_z}, \ref{Pmag_z} and \ref{rmsVz_z}.}
\label{runC1b}
\end{figure}

In order to investigate the issue of numerical convergence, we present the 
various profiles for run C1b, which is identical to run C1 but has
an effective resolution 2 times larger. As can be seen from Fig.~\ref{runC1b},
and Fig.~\ref{rho_z}, \ref{Pth_z}, \ref{Pkin_z}, \ref{Pmag_z}, \ref{rmsVz_z}
which displayed the results for run C1, 
 the  profiles 
present some moderate differences for the two cases implying that numerical convergence 
is not fully reached. In particular, the density profile is slightly less peaked 
for run C1b than for run C1. Similarly, the rms velocity is 
about 5 km s$^{-1}$ for run C1b while it is equal to about 4-4.5 km s$^{-1}$ for 
run C1. The pressures however present very comparable values and profiles 
between the two runs. 
Altogether, this shows that the quantities which characterize the disk structure 
are reasonably described at the resolution used in the paper.

\bibliographystyle{aa}
\bibliography{biblio}

\begin{thebibliography}{69}
\expandafter\ifx\csname natexlab\endcsname\relax\def\natexlab#1{#1}\fi

\bibitem[{{Agertz} {et~al.}(2013){Agertz}, {Kravtsov}, {Leitner}, \&
  {Gnedin}}]{agertz+2013}
{Agertz}, O., {Kravtsov}, A.~V., {Leitner}, S.~N., \& {Gnedin}, N.~Y. 2013,
  \apj, 770, 25

\bibitem[{{Alves} {et~al.}(2007){Alves}, {Lombardi}, \& {Lada}}]{alves+2007}
{Alves}, J., {Lombardi}, M., \& {Lada}, C.~J. 2007, \aap, 462, L17

\bibitem[{{Audit} \& {Hennebelle}(2005)}]{ah05}
{Audit}, E. \& {Hennebelle}, P. 2005, \aap, 433, 1

\bibitem[{{Audit} \& {Hennebelle}(2010)}]{ah+2010}
{Audit}, E. \& {Hennebelle}, P. 2010, \aap, 511, A76

\bibitem[{{Banerjee} {et~al.}(2009){Banerjee}, {V{\'a}zquez-Semadeni},
  {Hennebelle}, \& {Klessen}}]{banerjee+2009}
{Banerjee}, R., {V{\'a}zquez-Semadeni}, E., {Hennebelle}, P., \& {Klessen},
  R.~S. 2009, \mnras, 398, 1082

\bibitem[{{Blondin} {et~al.}(1998){Blondin}, {Wright}, {Borkowski}, \&
  {Reynolds}}]{blondin+1998}
{Blondin}, J.~M., {Wright}, E.~B., {Borkowski}, K.~J., \& {Reynolds}, S.~P.
  1998, \apj, 500, 342

\bibitem[{{Bournaud} {et~al.}(2010){Bournaud}, {Elmegreen}, {Teyssier},
  {Block}, \& {Puerari}}]{bournaud+2010}
{Bournaud}, F., {Elmegreen}, B.~G., {Teyssier}, R., {Block}, D.~L., \&
  {Puerari}, I. 2010, \mnras, 409, 1088

\bibitem[{{Chevalier}(1977)}]{chevalier77}
{Chevalier}, R.~A. 1977, \araa, 15, 175

\bibitem[{{Commer{\c c}on} {et~al.}(2011){Commer{\c c}on}, {Hennebelle}, \&
  {Henning}}]{commercon+2011}
{Commer{\c c}on}, B., {Hennebelle}, P., \& {Henning}, T. 2011, \apjl, 742, L9

\bibitem[{{Crutcher}(2012)}]{crutcher2012}
{Crutcher}, R.~M. 2012, \araa, 50, 29

\bibitem[{{Dale} {et~al.}(2012){Dale}, {Ercolano}, \& {Bonnell}}]{dale+2012}
{Dale}, J.~E., {Ercolano}, B., \& {Bonnell}, I.~A. 2012, \mnras, 424, 377

\bibitem[{{Dale} {et~al.}(2013){Dale}, {Ercolano}, \& {Bonnell}}]{dale+2013}
{Dale}, J.~E., {Ercolano}, B., \& {Bonnell}, I.~A. 2013, \mnras, 430, 234

\bibitem[{{de Avillez} \& {Breitschwerdt}(2005)}]{deavillez+2005}
{de Avillez}, M.~A. \& {Breitschwerdt}, D. 2005, \aap, 436, 585

\bibitem[{{Dib} {et~al.}(2006){Dib}, {Bell}, \& {Burkert}}]{dib+2006}
{Dib}, S., {Bell}, E., \& {Burkert}, A. 2006, \apj, 638, 797

\bibitem[{{Dib} {et~al.}(2010){Dib}, {Hennebelle}, {Pineda}, {Csengeri},
  {Bontemps}, {Audit}, \& {Goodman}}]{dib+2010}
{Dib}, S., {Hennebelle}, P., {Pineda}, J.~E., {et~al.} 2010, \apj, 723, 425

\bibitem[{{Dobbs} {et~al.}(2011){Dobbs}, {Burkert}, \& {Pringle}}]{dobbs+2011}
{Dobbs}, C.~L., {Burkert}, A., \& {Pringle}, J.~E. 2011, \mnras, 417, 1318

\bibitem[{{Dobbs} {et~al.}(2013){Dobbs}, {Krumholz}, {Ballesteros-Paredes},
  {Bolatto}, {Fukui}, {Heyer}, {Mac Low}, {Ostriker}, \&
  {V{\'a}zquez-Semadeni}}]{dobbs+2013}
{Dobbs}, C.~L., {Krumholz}, M.~R., {Ballesteros-Paredes}, J., {et~al.} 2013,
  ArXiv e-prints

\bibitem[{{Dubois} \& {Teyssier}(2008)}]{dubois+2008}
{Dubois}, Y. \& {Teyssier}, R. 2008, \aap, 477, 79

\bibitem[{{Faucher-Gigu{\`e}re} {et~al.}(2013){Faucher-Gigu{\`e}re},
  {Quataert}, \& {Hopkins}}]{faucher+2013}
{Faucher-Gigu{\`e}re}, C.-A., {Quataert}, E., \& {Hopkins}, P.~F. 2013, \mnras,
  433, 1970

\bibitem[{{Ferri{\`e}re}(2001)}]{ferriere+2001}
{Ferri{\`e}re}, K.~M. 2001, Reviews of Modern Physics, 73, 1031

\bibitem[{{Fromang} {et~al.}(2006){Fromang}, {Hennebelle}, \&
  {Teyssier}}]{fromang+2006}
{Fromang}, S., {Hennebelle}, P., \& {Teyssier}, R. 2006, \aap, 457, 371

\bibitem[{{Gazol} {et~al.}(2001){Gazol}, {V{\'a}zquez-Semadeni},
  {S{\'a}nchez-Salcedo}, \& {Scalo}}]{gazol+2001}
{Gazol}, A., {V{\'a}zquez-Semadeni}, E., {S{\'a}nchez-Salcedo}, F.~J., \&
  {Scalo}, J. 2001, \apjl, 557, L121

\bibitem[{{Gent} {et~al.}(2013){Gent}, {Shukurov}, {Fletcher}, {Sarson}, \&
  {Mantere}}]{gent+2013}
{Gent}, F.~A., {Shukurov}, A., {Fletcher}, A., {Sarson}, G.~R., \& {Mantere},
  M.~J. 2013, \mnras, 432, 1396

\bibitem[{{Heiles} \& {Troland}(2005)}]{heiles+2005}
{Heiles}, C. \& {Troland}, T.~H. 2005, \apj, 624, 773

\bibitem[{{Heitsch} {et~al.}(2008){Heitsch}, {Hartmann}, \&
  {Burkert}}]{heitsch+2008}
{Heitsch}, F., {Hartmann}, L.~W., \& {Burkert}, A. 2008, \apj, 683, 786

\bibitem[{{Hennebelle}(2013)}]{h2013}
{Hennebelle}, P. 2013, \aap, 556, A153

\bibitem[{{Hennebelle} \& {Audit}(2007)}]{ha07}
{Hennebelle}, P. \& {Audit}, E. 2007, \aap, 465, 431

\bibitem[{{Hennebelle} {et~al.}(2008){Hennebelle}, {Banerjee},
  {V{\'a}zquez-Semadeni}, {Klessen}, \& {Audit}}]{h+2008}
{Hennebelle}, P., {Banerjee}, R., {V{\'a}zquez-Semadeni}, E., {Klessen}, R.~S.,
  \& {Audit}, E. 2008, \aap, 486, L43

\bibitem[{{Hennebelle} \& {Chabrier}(2013)}]{hc2013}
{Hennebelle}, P. \& {Chabrier}, G. 2013, \apj, 770, 150

\bibitem[{{Hennebelle} {et~al.}(2011){Hennebelle}, {Commer{\c c}on}, {Joos},
  {Klessen}, {Krumholz}, {Tan}, \& {Teyssier}}]{h+2011}
{Hennebelle}, P., {Commer{\c c}on}, B., {Joos}, M., {et~al.} 2011, \aap, 528,
  A72

\bibitem[{{Hennebelle} \& {P{\'e}rault}(2000)}]{h+2000}
{Hennebelle}, P. \& {P{\'e}rault}, M. 2000, \aap, 359, 1124

\bibitem[{{Hill} {et~al.}(2012){Hill}, {Joung}, {Mac Low}, {Benjamin},
  {Haffner}, {Klingenberg}, \& {Waagan}}]{hill+2012}
{Hill}, A.~S., {Joung}, M.~R., {Mac Low}, M.-M., {et~al.} 2012, \apj, 750, 104

\bibitem[{{Hopkins} {et~al.}(2011){Hopkins}, {Quataert}, \&
  {Murray}}]{hopkins+2011}
{Hopkins}, P.~F., {Quataert}, E., \& {Murray}, N. 2011, \mnras, 417, 950

\bibitem[{{Inoue} \& {Inutsuka}(2012)}]{inoue+2012}
{Inoue}, T. \& {Inutsuka}, S.-i. 2012, \apj, 759, 35

\bibitem[{{Joung} \& {Mac Low}(2006)}]{joung+2006}
{Joung}, M.~K.~R. \& {Mac Low}, M.-M. 2006, \apj, 653, 1266

\bibitem[{{Joung} {et~al.}(2009){Joung}, {Mac Low}, \& {Bryan}}]{joung+2009}
{Joung}, M.~R., {Mac Low}, M.-M., \& {Bryan}, G.~L. 2009, \apj, 704, 137

\bibitem[{{Kim} {et~al.}(2011){Kim}, {Kim}, \& {Ostriker}}]{kim+2011}
{Kim}, C.-G., {Kim}, W.-T., \& {Ostriker}, E.~C. 2011, \apj, 743, 25

\bibitem[{{Kim} {et~al.}(2013){Kim}, {Ostriker}, \& {Kim}}]{kim+2013}
{Kim}, C.-G., {Ostriker}, E.~C., \& {Kim}, W.-T. 2013, \apj, 776, 1

\bibitem[{{Krumholz} {et~al.}(2004){Krumholz}, {McKee}, \&
  {Klein}}]{Krumholz+04}
{Krumholz}, M.~R., {McKee}, C.~F., \& {Klein}, R.~I. 2004, \apj, 611, 399

\bibitem[{{Kuijken} \& {Gilmore}(1989)}]{Kuijken+1989}
{Kuijken}, K. \& {Gilmore}, G. 1989, \mnras, 239, 571

\bibitem[{{Lazarian} \& {Vishniac}(1999)}]{lazarian+1999}
{Lazarian}, A. \& {Vishniac}, E.~T. 1999, \apj, 517, 700

\bibitem[{{Leroy} {et~al.}(2008){Leroy}, {Walter}, {Brinks}, {Bigiel}, {de
  Blok}, {Madore}, \& {Thornley}}]{leroy+2008}
{Leroy}, A.~K., {Walter}, F., {Brinks}, E., {et~al.} 2008, \aj, 136, 2782

\bibitem[{{Mac Low}(2013)}]{maclow2013}
{Mac Low}, M.-M. 2013, ArXiv e-prints

\bibitem[{{Mac Low} \& {Klessen}(2004)}]{maclow+2004}
{Mac Low}, M.-M. \& {Klessen}, R.~S. 2004, Reviews of Modern Physics, 76, 125

\bibitem[{{Matzner}(2002)}]{matzner2002}
{Matzner}, C.~D. 2002, \apj, 566, 302

\bibitem[{{McKee} \& {Ostriker}(1977)}]{mckee-ostriker1977}
{McKee}, C.~F. \& {Ostriker}, J.~P. 1977, \apj, 218, 148

\bibitem[{{Molina} {et~al.}(2012){Molina}, {Glover}, {Federrath}, \&
  {Klessen}}]{molina+2012}
{Molina}, F.~Z., {Glover}, S.~C.~O., {Federrath}, C., \& {Klessen}, R.~S. 2012,
  \mnras, 423, 2680

\bibitem[{{Myers} {et~al.}(2013){Myers}, {McKee}, {Cunningham}, {Klein}, \&
  {Krumholz}}]{myers+2013}
{Myers}, A.~T., {McKee}, C.~F., {Cunningham}, A.~J., {Klein}, R.~I., \&
  {Krumholz}, M.~R. 2013, \apj, 766, 97

\bibitem[{{Ostriker} {et~al.}(2010){Ostriker}, {McKee}, \&
  {Leroy}}]{ostriker+2010}
{Ostriker}, E.~C., {McKee}, C.~F., \& {Leroy}, A.~K. 2010, \apj, 721, 975

\bibitem[{{Padoan} \& {Nordlund}(2011)}]{padoan+2011}
{Padoan}, P. \& {Nordlund}, {\AA}. 2011, \apj, 730, 40

\bibitem[{{Pakmor} \& {Springel}(2013)}]{rudiger+2013}
{Pakmor}, R. \& {Springel}, V. 2013, \mnras, 432, 176

\bibitem[{{Passot} \& {V{\'a}zquez-Semadeni}(2003)}]{passot+2003}
{Passot}, T. \& {V{\'a}zquez-Semadeni}, E. 2003, \aap, 398, 845

\bibitem[{{Price} \& {Bate}(2008)}]{price+2008}
{Price}, D.~J. \& {Bate}, M.~R. 2008, \mnras, 385, 1820

\bibitem[{{Renaud} {et~al.}(2013){Renaud}, {Bournaud}, {Emsellem}, {Elmegreen},
  {Teyssier}, {Alves}, {Chapon}, {Combes}, {Dekel}, {Gabor}, {Hennebelle}, \&
  {Kraljic}}]{renaud+2013}
{Renaud}, F., {Bournaud}, F., {Emsellem}, E., {et~al.} 2013, \mnras, 436, 1836

\bibitem[{{Ryu} {et~al.}(2000){Ryu}, {Jones}, \& {Frank}}]{ryu+2000}
{Ryu}, D., {Jones}, T.~W., \& {Frank}, A. 2000, \apj, 545, 475

\bibitem[{{Santos-Lima} {et~al.}(2010){Santos-Lima}, {Lazarian}, {de Gouveia
  Dal Pino}, \& {Cho}}]{santos-lima+2010}
{Santos-Lima}, R., {Lazarian}, A., {de Gouveia Dal Pino}, E.~M., \& {Cho}, J.
  2010, \apj, 714, 442

\bibitem[{{Shu} {et~al.}(1987){Shu}, {Adams}, \& {Lizano}}]{shu+1987}
{Shu}, F.~H., {Adams}, F.~C., \& {Lizano}, S. 1987, \araa, 25, 23

\bibitem[{{Slyz} {et~al.}(2005){Slyz}, {Devriendt}, {Bryan}, \&
  {Silk}}]{slyz+2005}
{Slyz}, A.~D., {Devriendt}, J.~E.~G., {Bryan}, G., \& {Silk}, J. 2005, \mnras,
  356, 737

\bibitem[{{Sutherland} \& {Dopita}(1993)}]{sutherland+1993}
{Sutherland}, R.~S. \& {Dopita}, M.~A. 1993, \apjs, 88, 253

\bibitem[{{Tasker}(2011)}]{tasker2011}
{Tasker}, E.~J. 2011, \apj, 730, 11

\bibitem[{{Tasker} \& {Bryan}(2006)}]{tasker-bryan2006}
{Tasker}, E.~J. \& {Bryan}, G.~L. 2006, \apj, 641, 878

\bibitem[{{Teyssier}(2002)}]{teyssier2002}
{Teyssier}, R. 2002, \aap, 385, 337

\bibitem[{{Toomre}(1964)}]{toomre1964}
{Toomre}, A. 1964, \apj, 139, 1217

\bibitem[{{Troland} \& {Heiles}(1986)}]{troland+1986}
{Troland}, T.~H. \& {Heiles}, C. 1986, \apj, 301, 339

\bibitem[{{V{\'a}zquez-Semadeni} {et~al.}(2006){V{\'a}zquez-Semadeni}, {Ryu},
  {Passot}, {Gonz{\'a}lez}, \& {Gazol}}]{vazquez+2006}
{V{\'a}zquez-Semadeni}, E., {Ryu}, D., {Passot}, T., {Gonz{\'a}lez}, R.~F., \&
  {Gazol}, A. 2006, \apj, 643, 245

\bibitem[{{Veilleux} {et~al.}(2005){Veilleux}, {Cecil}, \&
  {Bland-Hawthorn}}]{veilleux+2005}
{Veilleux}, S., {Cecil}, G., \& {Bland-Hawthorn}, J. 2005, \araa, 43, 769

\bibitem[{{Wang} \& {Abel}(2009)}]{Wang+2009}
{Wang}, P. \& {Abel}, T. 2009, \apj, 696, 96

\bibitem[{{Wolfire} {et~al.}(2003){Wolfire}, {McKee}, {Hollenbach}, \&
  {Tielens}}]{wolfire+2003}
{Wolfire}, M.~G., {McKee}, C.~F., {Hollenbach}, D., \& {Tielens}, A.~G.~G.~M.
  2003, \apj, 587, 278

\bibitem[{{Zuckerman} \& {Evans}(1974)}]{zuckerman+1974}
{Zuckerman}, B. \& {Evans}, II, N.~J. 1974, \apjl, 192, L149

\end{thebibliography}

\end{document}